\documentclass{aa}  

\usepackage{graphicx}
\usepackage{txfonts}
\newcommand{\emy}{\epsilon(250 \,\mu\mathrm{m})}
\newcommand{\emyrat}{F(250 \,\mu\mathrm{m})/F(500 \,\mu\mathrm{m})}
\newcommand{\absH}{\sigma(250 \,\mu\mathrm{m})}
\newcommand{\absM}{\kappa(250 \,\mu\mathrm{m})}

\begin{document}

   \title{Dust emissivity and absorption cross section \\ in DustPedia\thanks{DustPedia is a collaborative focused research project supported by the European Union under the Seventh Framework Programme (2007- 2013) call (proposal no. 606824, P.I.\ J. I. Davies, {\tt http://www.dustpedia.com}). The DustPedia data is publicly available at {\tt http://dustpedia.astro.noa.gr}.} 
  late-type galaxies}


   \author{
S.~Bianchi\inst{1}
\and 
V.~Casasola\inst{2,1}
\and
M.~Baes\inst{3}
\and
C. J. R.~Clark\inst{4}
\and
E.~Corbelli\inst{1}
\and
J.~I.~Davies\inst{5}
\and
I.~De~Looze\inst{6,3}
\and
P.~De~Vis\inst{5}
\and
W.~Dobbels\inst{3}
\and
M.~Galametz\inst{7}
\and
F.~Galliano\inst{7}
\and
A.~P.~Jones\inst{8}
\and
S.~C.~Madden\inst{7}
\and
L.~Magrini\inst{1}
\and 
A.~Mosenkov\inst{9,10}
\and
A.~Nersesian\inst{11,12}
\and
S.~Viaene\inst{13,3}
\and
E.~M.~Xilouris\inst{11}
\and
N.~Ysard\inst{8}
}

\institute{
INAF - Osservatorio Astrofisico di Arcetri, Largo E. Fermi 5, I-50125, Florence, Italy\\
              \email{sbianchi@arcetri.astro.it}
\and
INAF - Istituto di Radioastronomia, Via P. Gobetti 101, I-40129, Bologna, Italy
\and
Sterrenkundig Observatorium, Universiteit Gent, Krijgslaan 281 S9, B-9000 Gent, Belgium
\and
Space Telescope Science Institute, 3700 San Martin Drive, Baltimore, Maryland, 21218, USA
\and
School of Physics and Astronomy, Cardiff University, The Parade, Cardiff CF24 3AA, UK 
\and
Department of Physics and Astronomy, University College London, Gower Street, London WC1E 6BT, UK
\and
AIM, CEA, CNRS, Universit\'e Paris-Saclay, Universit\'e Paris Diderot, Sorbonne Paris Cit\'e, F-91191 Gif-sur-Yvette, France
\and
Institut d'Astrophysique  Spatiale, CNRS, Universit\'e Paris Sud, Universit\'e Paris-Saclay, B\^at. 121,  F-91405 Orsay cedex, France
\and
Central Astronomical Observatory of RAS, Pulkovskoye Chaussee 65/1, 196140, St. Petersburg, Russia
\and
St. Petersburg State University, Universitetskij Pr. 28, 198504, St. Petersburg, Stary Peterhof, Russia
\and
National Observatory of Athens, Institute for Astronomy, Astrophysics, Space Applications and Remote Sensing,  Ioannou Metaxa and Vasileos Pavlou GR-15236, Athens, Greece
\and
Department of Astrophysics, Astronomy \& Mechanics, Faculty of Physics, University of Athens, Panepistimiopolis, GR-15784 Zografos, Athens, Greece
\and
Centre for Astrophysics Research, University of Hertfordshire, College Lane, Hatfield, AL10 9AB, UK
}

   \date{}

 
  \abstract
   {}
   {
  We compare the far-infrared to sub-millimetre dust emission properties measured in high Galactic latitude cirrus with those determined in a sample of 204 late-type DustPedia galaxies. The aim is to verify if it is appropriate to use Milky Way dust properties to derive dust masses in external galaxies.  
     }
   {
 We used {\em Herschel} observations and atomic and molecular gas masses to estimate $\emy$, the disc-averaged dust emissivity at 250 $\mu$m, and from this, the absorption cross section per H atom $\absH$ and per dust mass $\absM$. The emissivity $\emy$ requires one assumption, which is the CO-to-H$_2$ conversion factor, and the dust temperature is additionally required for $\absH$; yet another constraint on the dust-to-hydrogen ratio $D/H$, depending on metallicity, is required for $\absM$. 
    }
   {
We find $\emy = 0.82 \pm 0.07$ MJy sr$^{-1}$ ($10^{20}$ H cm$^{-2}$)$^{-1}$ for galaxies with $4 < \emyrat < 5$. This depends only weakly on the adopted  CO-to-H$_2$ conversion factor. 
The value is almost the same as that for the Milky Way at the same colour ratio. Instead, for $\emyrat > 6$, $\emy$ is lower than predicted by its dependence on the heating conditions. 
The reduction suggests a variation in dust emission properties for spirals of earlier type, higher metallicity,  and with a higher fraction of molecular gas.
When the standard emission properties of Galactic cirrus are used for these galaxies,
their dust masses  might be underestimated by up to a factor of two.
Values for $\absH$ and $\absM$ at the Milky Way metallicity are also close to those of the cirrus. Mild trends of the absorption cross sections with metallicity are found,
although the results depend on the assumptions made.
   }
   {}

              
\titlerunning{Dust emissivity and absorption cross section in late-type galaxies}
   \maketitle
%

\section{Introduction}

Dust grains of sub-micrometre size constitute a sizeable fraction of all the metals available in the interstellar medium (ISM). Their properties retain the imprint of several processes during the life of a galaxy:
grains are formed during star formation (in the atmospheres of giant stars and in the ejecta of supernovae). The grains together with gas are
destroyed as they coalesce into protostars. The grains
participate in the ISM evolution, when they accrete material from the gas phase or return to it after they are destroyed in
supernovae shocks. Knowing their total mass is therefore vital for understanding the chemical evolution of a galaxy \citep[for a recent review, see][]{GallianoARA&A2018}.

The dust mass is commonly derived by modelling the spectral energy distribution (SED) in the far-infrared (FIR) and sub-millimetre (submm) wavelength ranges \citep{HildebrandQJRAS1983}. The key ingredient in the estimate is knowing the wavelength-dependent FIR absorption 
(emission) cross section. This is typically derived by modelling the dust properties in the local ISM of the Milky Way (MW). A dust model consists of a mixture of (typically spherical) grains of different sizes and materials. Its composition is constrained by the metal depletion, that is, the difference between the (typically solar) elemental composition and that measured in the ISM gas phase. The missing metals have condensed into dust grains (for a review, see \citealt{DraineARA&A2003}).
Models were originally required to reproduce the dependence of extinction on wavelength (the extinction law) in 
the ultraviolet (UV), optical, and near-infrared (e.g. in the classical work of \citealt{MathisApJ1977}; additional constraints on polarisation are needed for models with aspherical grains, see e.g. \citealt{SiebenmorgenA&A2014}). 
With the increasing availability of observations at longer wavelengths from balloons and satellites, it has become possible to constrain the models in the FIR to submm.
This was done either by comparing the predicted FIR and submm absorption cross section with the cross sections that were estimated under a variety of assumptions in several locations in the Galaxy \citep{MezgerA&A1982,DraineApJ1984} or by simulating the emission due to
grains that were exposed to an interstellar radiation field (ISRF) and comparing this emission with the observed  emissivity, that is, the surface brightness per column density of hydrogen \citep{MezgerA&A1982,DesertA&A1990}.

In the past two decades, a common benchmark has been adopted at FIR to submm wavelengths. Dust models are required to reproduce the local MW emissivity, estimated from measurements of the high Galactic latitude cirrus from Far-InfraRed Absolute Spectrophotometer (FIRAS) data  \citep{WrightApJ1991,ReachApJ1995} or from Diffuse InfraRed Background Experiment (DIRBE) data \citep{ArendtApJ1998}. These two instruments operated on board the Cosmic Background Explorer \citep[COBE;][]{BoggessApJ1992}. The emissivity is predicted for grains that are exposed to the local ISRF (LISRF), estimated by \citet{MathisA&A1983}. 
Several dust models have been constructed and calibrated using these observational constraints
\citep{DwekApJ1997,LiApJ2001,DraineARA&A2003,ZubkoApJS2004,CompiegneA&A2011,JonesA&A2013,SiebenmorgenA&A2014,JonesA&A2017}. Despite the common starting point, the different choices for the dust materials, size distributions, and other constraints 
resulted in different absorption cross sections. As a result, dust mass estimates 
from FIR and submm observations
can vary by up to a factor $\sim$3 from model to model  
 \citep[see e.g.][]{SantiniA&A2014,CasasolaA&A2017,ChastenetA&A2017,NersesianA&A2019,HuntA&A2019}. 

In addition to the uncertainties of evaluating the emission properties of the MW cirrus,  variations in dust properties with the environment are also expected because grains 
can grow by accreting mantles of different compositions or by coagulating
in denser media, 
or because grains are processed in shocked environments (for the former, see \citealt{JonesA&A2017}; for the latter, \citealt{BocchioA&A2014}, and references in the two papers).
When the dust mass in external galaxies is estimated using lower resolution observations and global fluxes, there is no guarantee that the properties of the MW cirrus adequately represent the emission from the bulk of the dust. A solution is determining the dust absorption cross section {\em \textup{in situ}} by deriving the dust mass from the mass of metals that is available in the ISM of each galaxy, assuming a universal fraction of metals in dust and simplified dust heating conditions \citep{JamesMNRAS2002}. This method requires FIR observations, gas masses, and metallicities and has recently been reassessed by \citet[hereafter, C16]{ClarkMNRAS2016}, who found that the dust absorption cross sections in 22 high-metallicity objects are compatible with previous estimates within a large scatter, and that they did not {show large variations} from galaxy to galaxy. We here derive the
FIR dust emission properties following a similar approach. Based on data from the DustPedia project \citep{DaviesPASP2017}, our sample 
is an order of magnitude larger and spans a wider metallicity range than that of  C16.
We derive the emissivity, which is least dependent on assumptions and can be used as a benchmark for future dust models. We also derive the absorption cross section after further assumptions on the heating conditions and dust-to-gas ratio.

The paper is organised as follows: in Sects.~\ref{sec:sample} and \ref{sec:method} we present our sample and method.
Sect.~\ref{sec:mw} describes MW observations that we used for comparison with our results as well as the dust-to-gas ratio and metallicity estimates that are ingredients of our method.
Our results on the dust emissivity are presented and discussed in Sect.~\ref{sec:emy}. The results on the absorption cross section per H atom and per dust mass are reported in Sect.~\ref{sec:abs}.
We summarise the work and draw our conclusions in Sect.~\ref{sec:summa}.

\section{Sample and dataset}
\label{sec:sample}

The DustPedia sample \citep{DaviesPASP2017} includes 875 galaxies, almost all the large ($D_\mathrm{25} > 1\arcmin$) and nearby ($v < 3000$ km/s) objects that have been observed by the {\em Herschel} Space Observatory \citep{PilbrattA&A2010}.  For all these galaxies, photometric data are available in up to 34 bands from the UV to the submm \citep{ClarkA&A2018}. In the FIR and submm bands, we required that objects have available
 flux densities at 250 and 500 $\mu$m
from {\em Herschel}'s Spectral and Photometric Imaging Receiver \citep[SPIRE;][]{GriffinA&A2010}. We used fits to the full spectral energy distribution (SED) obtained by \citet{NersesianA&A2019} using the Code Investigating GALaxy Emission \citep[CIGALE;][]{BoquienA&A2019} coupled to The Heterogeneous dust Evolution Model for Interstellar Solids \citep[THEMIS;][]{JonesA&A2017}. 
We also used the procedures of \citet{NersesianA&A2019} to fit the SED for $\lambda\ge 100 \mu$m with a single-temperature modified black body (MBB).

The emissivity determination requires knowing the gas column density, therefore we conducted literature searches to collect information on the atomic and molecular gas. \ion{H}{i} masses are available for 87\% of the DustPedia  sample (\citealt{DeVisA&A2019}, Casasola et al., submitted). 
We searched for observations of the CO molecule (a tracer for H$_2$) in late-type galaxies 
(later than Sa, with Hubble stage $T\ge 0.5$) that were detected at 250 $\mu$m and found them for 255 galaxies (29\% of the full DustPedia  sample). H$_2$ masses were obtained from the CO observations assuming the CO-to-H$_2$ conversion factor of \citet{AmorinA&A2016}, which is dependent on the oxygen abundance  as $(O/H)^{-1.5}$. This is an intermediate choice between using a constant MW-based value \citep{BolattoARA&A2013} and stronger dependencies on metallicity \citep{HuntA&A2015b}. 
Because dust is well detected up to the optical radius $R_{25}=D_{25}/2$ \citep{PohlenA&A2010,CasasolaA&A2017}, we verified that estimates for the
gas refer to the same aperture. \ion{H}{i} observations are typically taken with beams larger than $R_{25}$, 
while most CO observations only cover the central region of a galaxy with a smaller beam. Using averaged radial profiles for each gas
component, we implemented aperture corrections to retrieve the mass of atomic  and molecular hydrogen  within $R_{25}$  
(for a full description, see Casasola et al., submitted).

Global O/H metallicities are available for about 60\% of the full DustPedia sample from a literature compilation and archival integral field unit observations \citep{DeVisA&A2019}. Among the various calibrations for the oxygen abundance given by \citet{DeVisA&A2019}, we chose the N2 method of \citet{PettiniMNRAS2004}, which is compatible with the conversion factor derived by  \citet{AmorinA&A2016} and compares well with direct electron-temperature-based determinations at both high and low metallicities \citep{CurtiMNRAS2017}.
The N2 metallicities have the advantage that they are available for a larger number of objects, because they require only a limited wavelength coverage of the spectrum.

In summary, we selected all $T\ge 0.5$ galaxies that were detected at 250 $\mu$m (except for a few objects whose fluxes were flagged because of an insufficient sky coverage), that were detected in atomic gas and in molecular gas, or that were detected in atomic gas with upper limits in molecular gas if the mass of the atomic component was twice higher than the error on the total gas estimate (19\% of the final selection; the complementary case of detection in CO with upper limits in \ion{H}{I} is not present in our database),
for which an N2 metallicity was available, and whose SED coverage was sufficient for an MBB fit. 
In total, the sample used in this work includes 204 objects.

\section{Method}
\label{sec:method}

In MW studies, the dust emissivity $\epsilon_\nu$  is defined as the surface brightness of dust emission $I_\nu$ per hydrogen column density $N_\mathrm{H}$. The normalisation provides information that is independent of the ISM density along the line of sight. For our galaxies, 
disc-averaged values for $I_\nu$ and $N_\mathrm{H}$ can be obtained from the integrated flux density of dust emission $F_\nu$ and 
the total hydrogen mass $M_\mathrm{H} = M_\ion{H}{i} + M_\mathrm{H_2}$. The emissivity can then be estimated as 
\begin{equation}
\epsilon_\nu = \frac{I_\nu}{N_\mathrm{H}} = \frac{\displaystyle \frac{F_\nu}{\Omega}}{\displaystyle  \frac{(M_\ion{H}{i} + M_\mathrm{H_2})}{m_\mathrm{H} \, \Omega \, d^2}},
\label{eq:emy}
\end{equation}
where $m_\mathrm{H}$ is the mass of the hydrogen atom. The gas masses were derived assuming a galaxy distance $d$, and the ratio $M_\mathrm{H}/d^2$ (as well as the  emissivity estimate) is thus independent of $d$. $\epsilon_\nu$ is also independent of the solid angle $ \Omega$ because the gas masses and dust flux density refer to the same sky area, that is, the optical disc within $R_{25}$. Eq.~1 should in principle include the contribution of \ion{H}{ii} to $N_\mathrm{H}$. However, this contribution is seldom considered in the MW and difficult to measure in external galaxies. Lacking the data to estimate it, we here neglect  \ion{H}{ii}. We discuss this point further in the next sections.

When all the dust grains in a galaxy are made of the same material, have the same size, and are exposed to the same ISRF, the dust emissivity can be written in the optically thin limit as
\begin{equation}
\epsilon_\nu = \sigma_\nu \, B_\nu(T_\mathrm{d}),
\label{eq:tau}
\end{equation}
where $\sigma_\nu$ is the absorption cross section per hydrogen atom and $B_\nu$ is the Planck function at the dust temperature $T_\mathrm{d}$  (we only considered emission at thermal equilibrium because stochastic heating is not dominant for the FIR regime). In the more realistic case of various grain sizes, materials, and heating conditions, $\sigma_\nu$  and 
$T_\mathrm{d}$ are different for each grain ($T_\mathrm{d}$ also depends on the  ISRF), and $\epsilon_\nu$ results from the integral of the second term of Eq.~\ref{eq:tau} 
over the dust and ISRF distributions. Nevertheless, the single-temperature simplification is widely used, together with a power-law description of the absorption cross section. At the reference wavelength of 250~$\mu$m, it is
\begin{equation}
\sigma_\nu = \sigma(250 \,\mu\mathrm{m}) \times \left(\frac{250 \,\mu\mathrm{m}}{\lambda}\right)^\beta.
\label{eq:tau_beta}
\end{equation}
Within this MBB approximation, Eq.~\ref{eq:tau} can be fitted to the observed emissivity SED of Eq.~\ref{eq:emy} to obtain an SED-averaged representation of $\sigma_\nu$. We discuss the effect of the single-temperature approximation and of the choice for the power-law index $\beta$ on the estimate of the absorption cross section later.

The absorption cross section per dust mass is
\begin{equation}
\kappa_\nu=\frac{\sigma_\nu}{m_\mathrm{H} \, {D/H}},
\label{eq:kappa}
\end{equation}
with $D/H$ the dust-to-hydrogen mass ratio. When the oxygen abundance in the gas ($O/H$) is a good proxy of the total metal abundance, and when a fixed fraction of metals is locked up in dust grains, $D/H$ should depend linearly on $O/H$ \citep[see, e.g.,][]{DraineApJ2007,MagriniA&A2011}. 
We  used
\begin{equation}
D/H = \left(\frac{D/H}{O/H}\right)_\mathrm{MW} \times O/H,
\label{eq:mwscale}
\end{equation}
scaling the ratio on the estimates of $D/H$ and $O/H$ for the MW. 
Because we used H-normalised quantities, we did not need to consider the contribution of helium (and other metals) to the gas mass.
In the appendix, we show that our approach is analogous to that used by \citet{JamesMNRAS2002} and C16 to derive $\kappa_\nu$.

Using Eq.~\ref{eq:tau} to~\ref{eq:kappa}, {we can write}
\begin{equation}
\emy = m_\mathrm{H} \,\, {D/H} \,\, \absM \, B_{\nu = c/250 \mu\mathrm{m}}(T_\mathrm{d}),
\label{eq:tau250}
\end{equation}
which clearly shows the expected dependence of the emissivity under the assumptions made here on the 
dust-to-hydrogen ratio, absorption cross section per dust mass, and 
dust temperature.

\begin{figure*}
\includegraphics[width=\hsize]{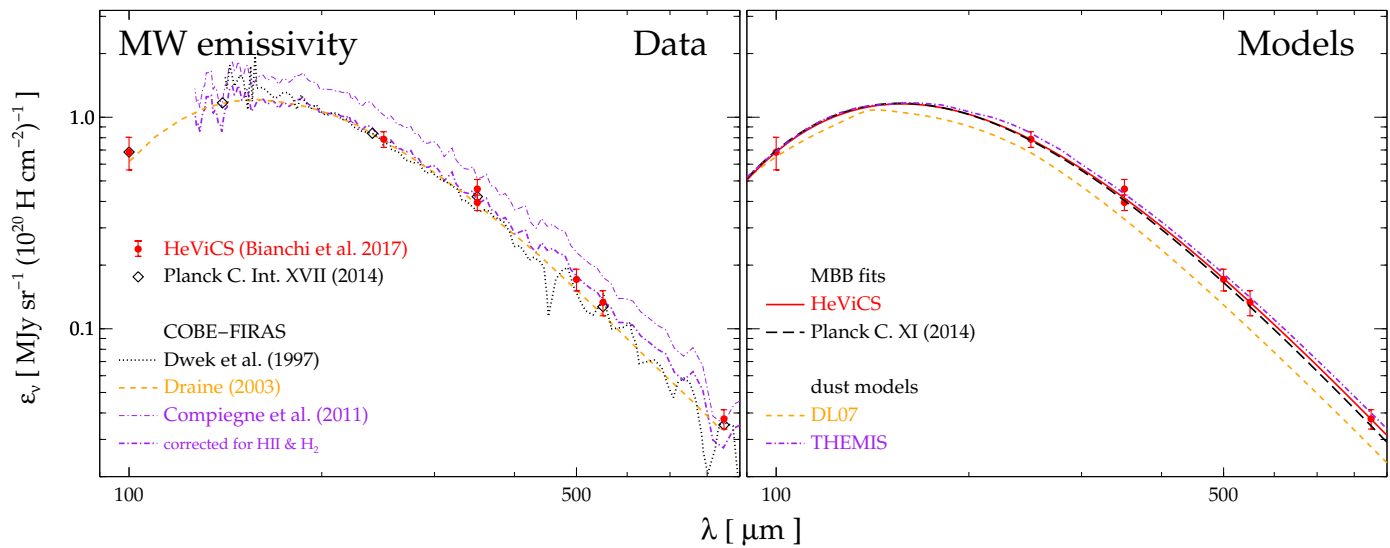}
\caption{
Milky Way emissivity $\epsilon_\nu$ at high Galactic latitude. In the left panel we compare estimates obtained from COBE-DIRBE and {\em Planck} data of the south Galactic pole \citep{PlanckIntermediateXVII} and from IRAS, {\em Herschel,} and {\em Planck} data of the HeViCS fields \citep{BianchiA&A2017} to those derived from the COBE-FIRAS spectrum. In the right panel, we show MBB fits to the HeViCS data \citep{BianchiA&A2017}  and to IRAS and {\em Planck} data over the whole diffuse high-Galactic latitude cirrus \citep{Planck2013XI}, together with the predictions from the THEMIS and \citet{DraineApJ2007} dust models heated by the LISRF. The HeViCS data points are repeated to facilitate comparison. 
}
\label{fig:emysed}
\end{figure*}
\begin{table*}
\caption{Milky Way reference values adopted here (in bold) and other values from the literature.}              
\label{tab:mwrefs}      
\small
\centering                                      
\begin{tabular}{c c l  l c}          
\hline\hline                        
Quantity & \multicolumn{2}{c}{value}  & notes & refs \\
\hline                                   
\rule{0pt}{3ex} 
$\emy$ & $\mathbf{ 0.79 \pm 0.04}$ & MJy sr$^{-1}$ ($10^{20}$ H cm$^{-2}$)$^{-1}$ &  & 1 \\
\rule{0pt}{3ex} 
$\emyrat$ &  $\mathbf{ 4.6 \pm 0.3}$ & &  & 1 \\
\rule{0pt}{3ex} 
$\absH$    &  $\mathbf{ 0.52 \pm 0.05}$ & $10^{-25}$ cm$^2$ H$^{-1}$ & $\beta=1.6$; $T_\mathrm{d} = 20.0 \pm 0.7$ K &  1\\
                  & $0.49 \pm 0.13$  &       &$\beta=1.59\pm0.12$; $T_\mathrm{d} = 20.3\pm1.3$ K & 2 \\
                  & $0.55 \pm 0.05$  &       &$\beta=1.65\pm0.10$; $T_\mathrm{d} = 19.8\pm1.0$ K & 3  \\
\rule{0pt}{3ex} 
$\absM$    & $\mathbf{ 3.7 \pm 0.7}$ & $\mathrm{cm}^2 \;\mathrm{g}^{-1}$ & using $D/H$ below& \\
                  & 6.4  & & THEMIS model ($\beta=1.79$) &4 \\
                  & 4.0  & & DL07      model ($\beta=2.08$) &5\rule[-1.2ex]{0pt}{0pt}\\

\hline  
\rule{0pt}{3ex} 
$D/H$        & $\mathbf{0.0085\pm0.0015}$ &  &         &  \\
                  & $0.0074-0.0093$    &        &depletions for $F_\star=0.4-0.8$                 & 6 \\
                  & $0.009-0.01$          &        &depletions for $\zeta$ Oph ($F_\star=1$)    & 6 \\
                  & $0.0074      $          &        &THEMIS dust model                                     & 7 \\
                  & $0.0104      $          &        &DL07 dust model                                                    &8 \\   
                  & $0.008-0.009$        &        &dust model                                                    &9 \\
\rule{0pt}{3ex} 
$12+\log(O/H)$ & $\mathbf{8.6\pm0.1}$ &       &  \\
                          & $8.5-8.6$        &        &absorption lines to nearby stars                     &10,11\\
                          & $8.59$            &        &depletions for nearby stars                            &12\\
                          & $8.66$            &        &depletions for $F_\star=0.4$                          &6 \\
                          & $8.5$              &        &depletions for $F_\star=1.0$                          &13,14 \\
                          & $8.5$              &        &\ion{H}{ii} region emission lines                    &15,16\rule[-1.2ex]{0pt}{0pt}\\
\hline                                             
\end{tabular}

\tablebib{(1)~\citet{BianchiA&A2017};
(2)~\citet{Planck2013XI};
(3)~\citet{PlanckIntermediateXVII};
(4)~\citet{GallianoARA&A2018};
(5)~\citet{BianchiA&A2013};
(6)~\citet{DraineBook2011};
(7)~\citet{JonesA&A2017};
(8)~\citet{DraineApJ2007};
(9)~\citet{ZubkoApJS2004};
(10)~\citet{MeyerApJ1998};
(11)~\citet{JensenApJ2005};
(12)~\citet{PrzybillaApJ2008};
(13)~\citet{JenkinsApJ2009};
(14)~\citet{RitcheyApJS2018};
(15)~\citet{PilyuginA&A2003};
(16)~\citet{EstebanMNRAS2018}.
}
\end{table*}

\section{MW reference values}
\label{sec:mw}

We used the MW cirrus emissivity obtained by \citet{BianchiA&A2017} from  {\em Herschel} Virgo Cluster Survey data \citep[HeViCS;][]{DaviesA&A2010}. The quantity was derived in all SPIRE bands. It can therefore be directly compared with estimates that are available for all the galaxies considered here. The HeViCS emissivity SED, which also uses  IRAS and {\em Planck} data  (Fig.~\ref{fig:emysed}, left panel), is in excellent
agreement with {\em Planck} determinations over much larger sky areas \citep{PlanckIntermediateXVII,Planck2013XI}. We used
$\emy$ averaged over the whole HeViCS field and the flux (emissivity) ratio $\emyrat$ to characterise the SED shape. The numerical values
of these quantities and of others presented in this section are given in Table~\ref{tab:mwrefs}. \citet{BianchiA&A2017} furthermore derived  $\absH$ by fitting the SED with  $\beta=1.6$ (\citealt{Planck2013XI}; Fig.~\ref{fig:emysed}, right panel).

The HeViCS and Planck Collaboration SEDs agree with FIRAS determinations \citep{DwekApJ1997,DraineARA&A2003}. A notable exception is the FIRAS spectrum for the high-latitude cirrus in \citeauthor{CompiegneA&A2011} (\citeyear{CompiegneA&A2011}; thin purple dot-dashed line in the left panel of Fig.~\ref{fig:emysed}). These cirrus estimates only consider the \ion{H}{i} contribution to
$N_\mathrm{H}$ and neglect \ion{H}{ii} (the contribution of high-latitude molecular gas is very low, a few percent at most; \citealt{CompiegneA&A2011}). 
Curiously, when \citet{CompiegneA&A2011} included a correction for other contributions to $N_\mathrm{H}$ (mostly a 20\% increase {due to \ion{H}{ii}}; see also \citealt{DraineBook2011}), their FIRAS emissivity becomes very similar to the other estimates (thick purple dot-dashed line in the left panel of Fig.~\ref{fig:emysed}). The discrepancy might be due to  different  methods, \ion{H}{i} observations, and determination of zero-levels in the derivation of the emissivity\footnote{\citet{BianchiA&A2018} erroneously attributed the discrepancy to a bias in the calibration of FIRAS data, which might only be significant around the 350 $\mu$m {\em Planck} and {\em Herschel} bands \citep{Planck2013VIII}, however.}.

In Fig.~\ref{fig:emysed} (right panel) we  show a few model predictions for dust grains heated 
by the LISRF\footnote{Emission from the THEMIS model has been computed with the DustEM code 
\citep{CompiegneA&A2011}. The \citet{DraineApJ2007b} emission templates are 
available at {\tt https://www.astro.princeton.edu/$\sim$draine/}.}. The THEMIS model (purple dot-dashed line) is able to reproduce the MW SED.
Because the model was optimised to reproduce the corrected FIRAS spectrum of 
\citet{CompiegneA&A2011}, it passes through the many other uncorrected estimates we
just discussed. Thus, its grain properties have to be considered as scaled on the non-ionised
hydrogen alone. The \citet{DraineApJ2007b} model (DL07 hereafter; orange dashed line in the figure) was instead originally set to match FIRAS data  \citep[see e.g.][]{LiApJ2001,DraineARA&A2003}, but it has since been rescaled and its predictions 
at 250 $\mu $m are a factor 0.65 lower than observations \citep[see also][]{BianchiA&A2013,PlanckIntermediateXVII} .

\begin{figure*}
\sidecaption
\includegraphics[width=12cm,clip=true,trim=0 0 0 0]{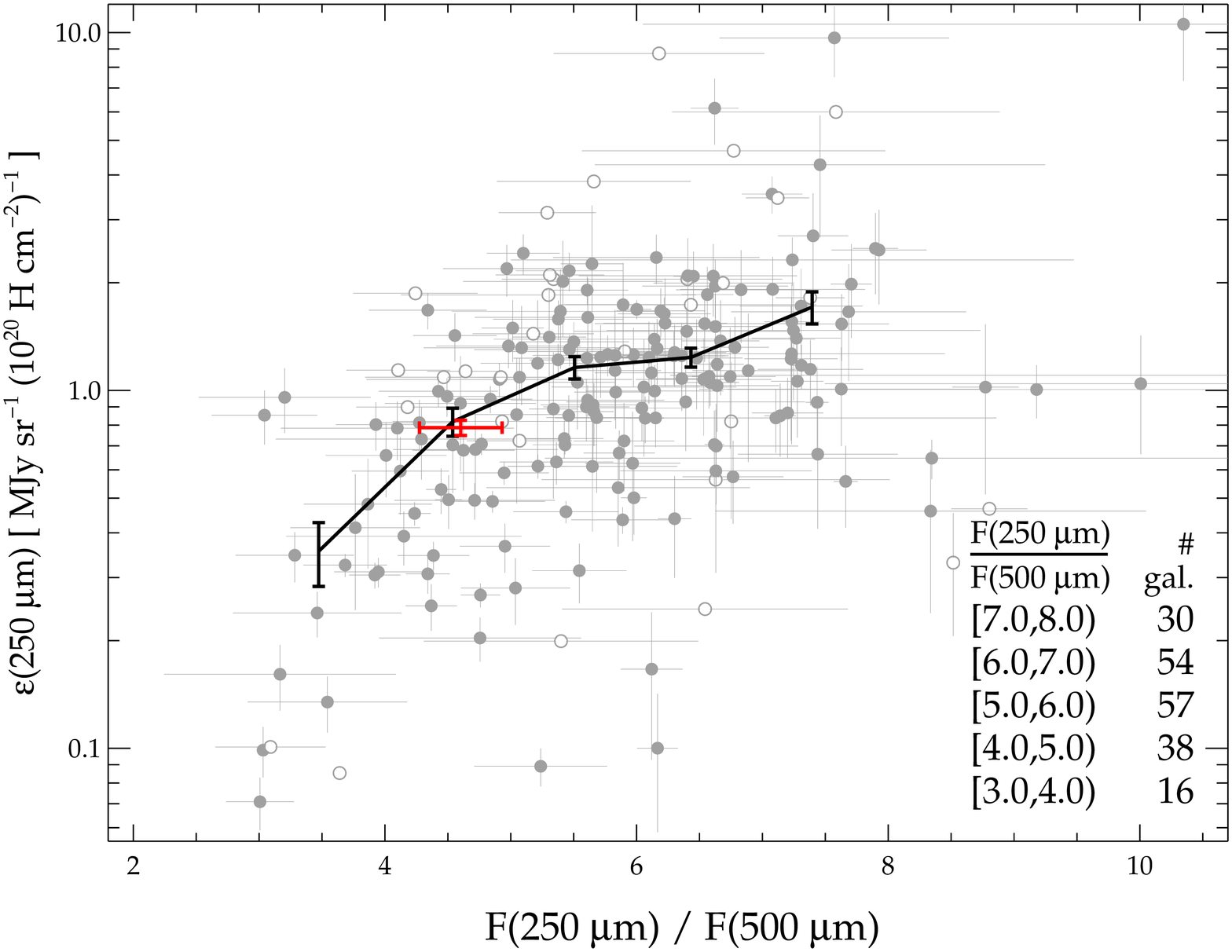}
\caption{
Emissivity $\emy$ as a function of the $\emyrat$ ratio. When data on the abscissa (and/or ordinate) are estimated at a level  below 2$\sigma$, the x (and/or y) error is omitted and an open symbol is used. The red error bar shows the MW estimate from HeVICS. 
The mean of $\emy$ and its standard deviation for five bins in $\emyrat$ is shown by the black error bars (connected by a solid line). The $\emyrat$ range and the number of galaxies is given for each bin.}
\label{fig:emy_vs_frat1}
\end{figure*}

To scale Eq.~\ref{eq:mwscale}, we need estimates of $D/H$ and $O/H$ for the same local medium for which 
the emissivity is derived. Following \citet{DraineApJ2007}, we derived $(D/H)_\mathrm{MW}$  from the difference between 
the total metal abundances of the interstellar medium (typically represented by the solar abundances or those of the pre-solar nebula) 
and the gas-phase abundances measured from absorption line spectra to nearby stars: the missing atoms are depleted into dust. \citet{JenkinsApJ2009} studied the element abundance along several lines of sight in the local Galaxy and parametrised the amount 
of atoms that are missing from the gas phase with a factor $F_\star$. We estimated $D/H$ for lines of sight of moderate to high 
depletion ($F_\star = $ 0.4 to 1, using Tables 9.5 and 23.1 in \citealt{DraineBook2011}). Other values for $D/H$ can be obtained 
from MW dust grain models. The adopted $(D/H)_\mathrm{MW}$  and uncertainty encompasses these values.
In the same way, we chose $(O/H)_\mathrm{MW}$ using metallicity measurements from absorption lines of nearby stars and
emission lines in  local \ion{H}{ii} regions (see Table~\ref{tab:mwrefs}, which also provides $\absM$ obtained from Eq.~\ref{eq:kappa}). 
With the adopted values, it is $((D/H)/(O/H))_\mathrm{MW} = 21 \pm 6$. 

\section{Emissivity}
\label{sec:emy}

\begin{figure*}
\includegraphics[width=\hsize]{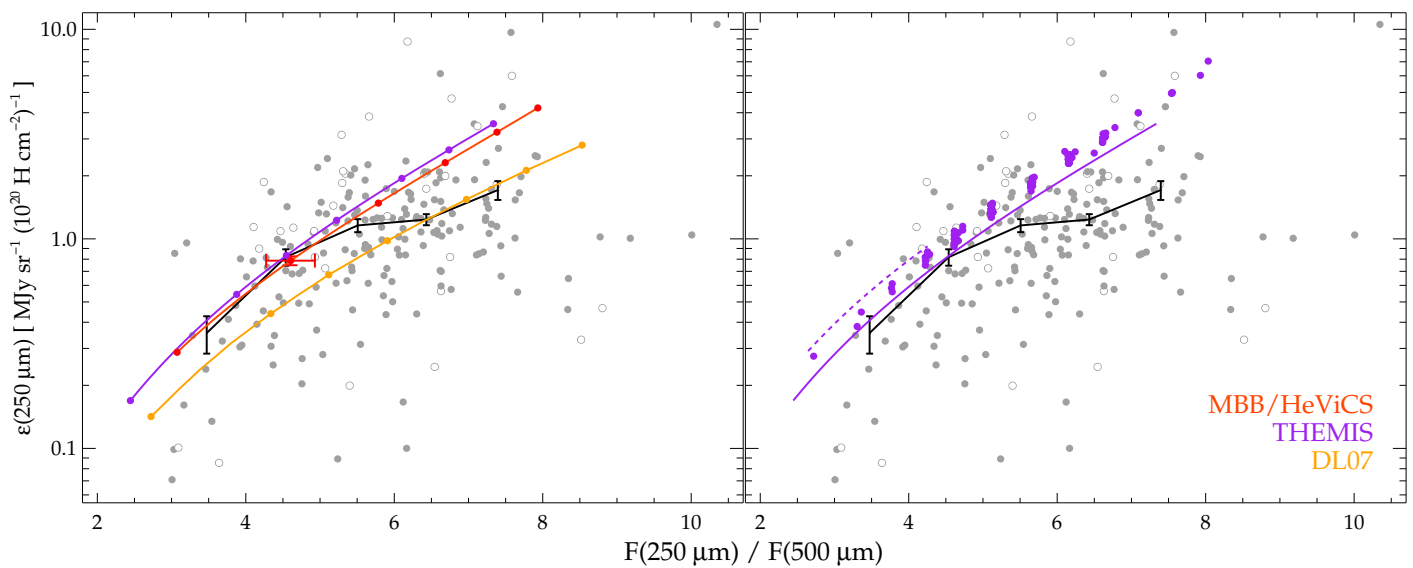}
\caption{
Same as Fig.~\ref{fig:emy_vs_frat1} (individual error bars are omitted for clarity). In the left panel, the red line shows the locus of an MBB with $\beta=1.6$ for different temperatures (red dots along the line give the values at $T_\mathrm{d} =$ 15, 20, 25, 30, 35, and 40 K, from left to right).
The purple and orange lines show the loci of dust heated by the LISRF scaled by a factor $U$ for the THEMIS and DL07 grain models, respectively (dots along the lines are for $U$ = 0.1, 0.5, 1, 2, 5, 10, and 20, from left to right). In the right panel, the THEMIS dust model (solid purple line) is repeated together with the CMM model from \citet{KoehlerA&A2015} for $0.1 < U <0.5$ (dashed purple line). Purple dots are the emissivity values estimated from the CIGALE-THEMIS best-fit SEDs of each galaxy (see text for details).}
\label{fig:emy_vs_frat2}
\end{figure*}

The emissivities at $250\;\mu\mathrm{m}$ derived for our DustPedia galaxies are shown in Figs.~\ref{fig:emy_vs_frat1} to~\ref{fig:emy_vs_frat2} as a function of the SPIRE 
$\emyrat$ ratio. This FIR colour is available for all galaxies in the sample; it is sensitive to the dust property variation with wavelength
(and to a lesser extent to the intensity of the radiation field; \citealt{SmithMNRAS2019}).
By studying $\emy$, we wish to provide a reference to benchmark dust models that depend least on assumptions. The estimate of $\emy$ using Eq.~\ref{eq:emy} does not require a model for the heating conditions, nor a recipe for the $D/H$ variations across the sample. Still, there is a major uncertainty in $\emy$: the uncertainty on the CO-to-H$_2$ conversion factor. 

\subsection{Results}
\label{sec:emyres}

Despite the large scatter of the data points, Fig.~\ref{fig:emy_vs_frat1} shows a mild trend, with a larger $\emy$ for bluer $\emyrat$ ratios. The Kendall correlation coefficient is $\tau_\mathrm{K}=0.30$ (with a negligible probability for the null hypothesis, $p_\mathrm{K}=0$). Although the different properties of dust in various objects might contribute to the scatter, most of it is compatible with the uncertainties in the measurements. In particular, the mean relative error of the emissivity is 28\%, dominated by the error on $N_\mathrm{H}$ (including 
estimates for the uncertainties in the conversion factor and aperture corrections), while it is 8\% for the colour ratio.

The average trend is shown in Fig.~\ref{fig:emy_vs_frat1}, where we  plot the mean and its standard deviation for five bins in $\emyrat$ (black error bars and solid line; all means presented in this work are clipped at 4$\sigma$ to exclude the most extreme outliers, which range
from none to four objects at most). The value for the $4 < \emyrat < 5$ bin
is entirely consistent with that for the MW cirrus at the same colour ratio (red data point; the numerical value is given in Table~\ref{tab:results}). For galaxies with higher colour ratios, 
$\emyrat \ga 6$, an apparent change in the trend is visible. 

\begin{figure}
\center
\includegraphics[width=8.5cm]{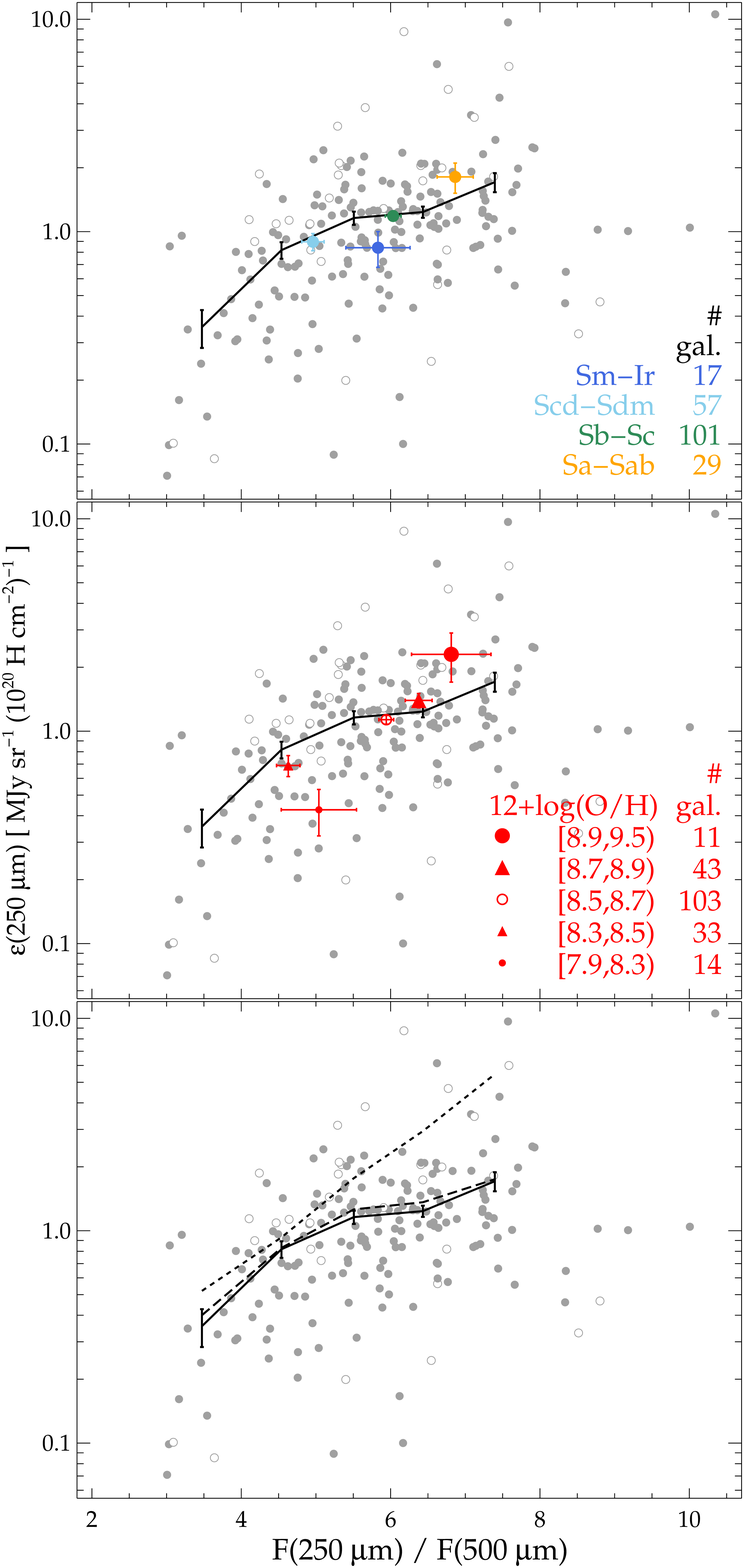}
\caption{
Emissivity $\emy$ as a function of $\emyrat$ for each galaxy (grey datapoints) and for the mean for $\emyrat$ bins
(same as Fig.~\ref{fig:emy_vs_frat1}, but individual error bars are omitted for clarity). 
Coloured dots in the top panel show the mean and its standard deviation for bins of different morphological types. 
Red dots in the central panel show the same, but for bins in metallicity (the number of objects for each bin is given in both panels). 
In the bottom panel, the long-dashed line shows the binned mean assuming a constant CO-to-H$_2$ conversion factor, and the short dotted line shows the 
mean when the H$_2$ component is neglected.
}
\label{fig:emy_vs_frat1a}
\end{figure}

Even though our sample is dominated by Sb-Sc galaxies (about half of the objects), a mild dependence of the emissivity on morphology is present. This can be seen in Fig.~\ref{fig:emy_vs_frat1a}  (top panel), where the mean and its standard deviation are plotted for 
each of four morphology bins (as defined in \citealt{BianchiA&A2018}). The trend is along that with colour ratio, emissivity progressively decreases from Sa-Sab (green) to Scd-Sm (blue). Only the Sm-Ir galaxies (dark blue) do not follow the sequence and fall  in between the other types. This behaviour reflects the average FIR colour of each morphology bin: the "peak" temperature of the average SED decreases from Sa-Sab  to Scd-Sm, while for Sm-Ir,  it is the same as that of Sb-Sc \citep{BianchiA&A2018}.

The emissivity is also found to mildly correlate with metallicity ($\tau_\mathrm{K}=0.31$, $p_\mathrm{K}=0$; not shown).
In addition to the dependence of the dust properties on elemental abundances, the correlation might again be the reflection of a 
trend of metallicity with colour ratio ($\tau_\mathrm{K}=0.34$, $p_\mathrm{K}=0$; not shown), which in turn might come from a 
concatenation of relations:  it is well known that the metallicity correlates with stellar mass \citep[the mass-metallicity relation;][]{TremontiApJ2004}; in later type spirals of higher stellar mass, a larger radiation fraction is absorbed by dust and is re-emitted in the FIR \citep{BianchiA&A2018}. The FIR luminosity correlates with the dust temperature, and thus with the colour ratio \citep{SymeonidisMNRAS2013,MagnelliA&A2014}.
In Fig.~\ref{fig:emy_vs_frat1a} (central panel) we show the emissivity versus colour ratio binned according to the metallicity. The metallicity in our sample ranges from 12 + $\log(O/H)$ = 7.9 to 9.5. The vast majority of objects has $8.3 \le 12 + \log(O/H) < 8.9$: the three bins in this range align with the emissivity trend with colour ratio (and morphology). The mean for objects in the lowest (and highest) metallicity bin is below (and above) the trend, although the scatter of the objects in these subsample is large (the higher metallicities are also beyond the range of applicability of the N2 calibration; \citealt{PettiniMNRAS2004}). We also explored the variations of $\emy$ with other galactic properties derived by \citet{NersesianA&A2019}, such as the bolometric luminosity, the contribution of young and old stars to it, the stellar mass ($M_\star$), the global and specific star formation rate ($\mathit{sSFR}$). We found no other  correlation stronger than those with colour ratio and metallicity (see below).

Because the metallicities of most of the objects are evenly distributed around the MW value (and close to it), the metallicity-dependent  CO-to-H$_2$ conversion factor we adopted does not change greatly throughout our sample. As a result, the emissivities we derived do not depend significantly on the conversion: for a constant Galactic value \citep[e.g. the value recommended by][]{BolattoARA&A2013}, the trend with colour ratio and the absolute value of the emissivity are very similar and virtually indistinguishable within the large uncertainty of the data. This trend is shown by the long-dashed line in Fig.~\ref{fig:emy_vs_frat1a} (bottom panel). The short-dashed line in the same figure instead shows the effects of neglecting the molecular gas component in the derivation of $\emy$ from Eq.~\ref{eq:emy}. The difference with the full derivation 
(solid line) highlights the galaxies for which H$_2$ is a significant contributor to the gas column density: for the higher colour ratios, 
$\emyrat \ga 6$, $M_\mathrm{H_2}/M_\ion{H}{i} \approx$ 1 to 2. A similar result, with a larger molecular gas fraction for galaxies with a bluer FIR colour, was found by \citet{GrovesApJ2015}.

Galaxies of type Sb-Sc and those with $8.5 \le 12 + \log(O/H) < 8.7$ have $\emy\approx$ 1.2 MJy sr$^{-1}$ ($10^{20}$ H cm$^{-2}$)$^{-1}$  at  $\emyrat \approx 6$ (see top and middle panel in Fig~\ref{fig:emy_vs_frat1a}).
When the MW is considered as an object of similar morphology \citep{HodgePASP1983,VanDerKruitProc1990} and metallicity, its average $\emy$ is a factor 1.5 higher than that estimated on the MW cirrus.
The difference, however, might be due in part to the warmer dust and in part to a larger contribution of molecular gas in the Galaxy
as a whole than in the local high-latitude environment we used as a reference.

\subsection{Discussion}
\label{sec:emi_d}

In Fig.~\ref{fig:emy_vs_frat2} (left panel) we compare the observed trends with the predictions from the MBB approximation and the THEMIS and DL07 dust grain models. In all cases, the SPIRE fluxes were derived by integrating the SEDs over the instrument 
filter response functions (The SPIRE Handbook, v. 2.5, 2014). The red line shows the trend of MBB emission with temperature,
assuming $\beta=1.6$ (the MW cirrus value).  The increase in $\emyrat$ across most of the range of
our sample can be reproduced by dust temperatures raising from  $T_\mathrm{d}  = 15$~K to 40~K
($T_\mathrm{d} \approx 20$ K is the value for the MW cirrus). 

The trends for dust models were derived assuming that grains are heated by the LISRF reported by \citet{MathisA&A1983} scaled by a factor $U$. 
The purple and orange lines in Fig.~\ref{fig:emy_vs_frat2} (left panel) show the results for the THEMIS and  DL07 models,
respectively. When the intensity of the radiation field through $U$ increased (we used values from 0.1 to 20), the model emissivity rises 
in the same fashion as the MBB does with $T_\mathrm{d}$. As expected from Fig.~\ref{fig:emysed} (see also Sect.~\ref{sec:mw}), the THEMIS 
emissivity for $U=1$ passes close to the MW estimate and its trend is very similar to that of the MBB, while the DL07 emissivity is lower and with a higher $\emyrat$ ratio (as can be seen by comparing the $U=1$ dots in this model  and in THEMIS). 
For $\emyrat \la 5.5$, the average trend follows the modelled MW-scaled MBB (and THEMIS). For objects with {larger $\emyrat$ }, the average emissivity is instead a factor $\approx 0.5-0.6$ of the MBB predictions (while it is fortuitously matched by DL07). {For the bin of the largest $\emyrat$},
the MBB is 9$\sigma$ away from the sample average. 

Apparently, the lower emissivity  is found for galaxies with a warmer colour ratio and a larger contribution of molecular hydrogen to $N_\mathrm{H}$ (Sect.~\ref{sec:emyres}). This is at odds with theoretical expectations because the grain emissivity should increase in denser environments. For example, \citet{KoehlerA&A2015} studied the variations of the THEMIS models in denser media, as the grains accrete layers of aliphatic-rich  amorphous carbon, coagulate, and grow ice-mantles. In all these stages the dust emissivity increase for the same $U$. The changes are expected to occur in regions where UV radiation is strongly attenuated and the ISRF is lower. As a reference, 
we plot in Fig.~\ref{fig:emy_vs_frat2} (right panel) the emissivity for $0.1<U<0.5$ from the CMM model of \citealt{KoehlerA&A2015}. In this and other models in that work, the emissivity is higher than 
the corresponding value from the THEMIS model for diffuse dust.

However, it is not likely that these dense regions contribute significantly to the 
global FIR emission. In observations at low spatial resolution, the SED could be 
biased against the emission from cold dust and in favour of grains at hotter temperature  (\citealt{GallianoARA&A2018} call it "a Matriochka effect"; see also \citealt{UtomoApJ2019}).
The emission from regions that contain colder grains, such as those where the emissivity could be 
enhanced, might contribute little to the global SED, in particular if the filling factor is low.
The bias could be even stronger in our work, where we used global fluxes.
On the other hand, variation in dust properties is also expected in regions with stronger 
radiation fields \citep{JonesA&A2017}. 
It might be wondered whether the dust evolution in this regime might cause the reduced dust emissivity in galaxies with warmer dust. 
Following Eq.~\ref{eq:tau250} and assuming no variations in the dust temperature and $D/H$, a smaller
$\emy$ can be obtained by reducing $\absM$. Thus, in galaxies with bluer FIR colours with 
reduced emissivity, the dust masses could be about a factor of two higher than those obtained with the 
standard properties of the diffuse MW dust. In the companion paper to this work, Casasola et al. (submitted)
find  lower $D/H$ ratios for objects with a larger H$_2$ contribution (that correspond
to galaxies that are bluer in the FIR, as shown in Sect.~\ref{sec:emyres} and in the bottom panel of Fig.~\ref{fig:emy_vs_frat1a}).
The explanation may be that the dust mass is underestimated when the THEMIS diffuse dust
properties are used for these objects. However, in a dust evolutionary scenario the changes in $D/H$, dust heating,
and absorption cross section are probably intertwined. For example, \citet{ChastenetA&A2017} fit the
resolved infrared emission in the low-metallicity Large and Small Magellanic Clouds under various assumptions for the 
dust heating by changing the relative proportion of the material components of THEMIS.
The resulting models have higher $\absM$ and a lower $D/H$ than the original model.  
The galaxies with  $\emyrat \ga 6$ typically have higher metallicities (Sect.~\ref{sec:emyres} and the middle panel 
of Fig.~\ref{fig:emy_vs_frat1a}) than the Magellanic Clouds, and thus a reduction in $D/H$ is less likely than a reduction in
$\absM$. A full dust modelling is needed to conclude, however.

As we discussed in the previous paragraph, the emissivity we derived from the global fluxes might be biased by different heating conditions within each galaxy.  In order to estimate the effect of this temperature mixing,
we derived mock emissivities from the THEMIS-based CIGALE best-fit SEDs of DustPedia galaxies \citep{NersesianA&A2019}. 
The fitting model,
following the the approach of \citet{DraineApJ2007}, assumes that part of the dust is heated by a radiation field characterised
by $U_\mathrm{min}$, and another part is heated by a power-law distribution of the heating fields, $U^{-\alpha}$, with $U\ge U_\mathrm{min}$ and 
$\alpha=2$. Typically, a fit results in most of the dust heated by the $U_\mathrm{min}$ component (still including temperature variations depending on the grain size and composition). A small fraction (up to a few percent at most) is instead needed to
account for the SED at wavelengths smaller than the thermal peak \citep[see e.g. ][]{DraineApJ2007,DaleApJ2012,HuntA&A2019,NersesianA&A2019}. 
Basically, the SED fitting operates by scaling the $N_\mathrm{H}$-normalised spectra produced by THEMIS to the true fluxes $F_\nu$. The normalisation yields the mass of gas $M_\mathrm{H}$ that is required for the dust emission, from which the dust mass $M_\mathrm{d}$ is derived 
for the (single-value) $D/H$ ratio of the model.
By reversing the process, we used the best-fit flux densities and gas masses (from $M_\mathrm{d}$) to derive $\emy$ and $\emyrat$ for each of the galaxies in our sample, using Eq.~\ref{eq:emy}.
The results of the mock derivation of $\emy$ versus $\emyrat$ are shown by the purple dots in Fig.~\ref{fig:emy_vs_frat2}. 

The CIGALE-THEMIS models fit the SED. This means that the mock $\emyrat$ spans a similar range of values as the observed ones. 
The mock estimate of  $\emy$ instead relies on the validity of the conversion of the gas mass into a dust mass through the 
hydrogen-scaled properties of the dust grain model. 
The mock $\emy$ is slightly higher than the THEMIS model for the MW 
(solid purple line) because in the CIGALE fits the fraction of small hydrocarbon solids was allowed to vary with respect to 
the standard value for diffuse dust. It was found that DustPedia galaxies are typically better fitted by a smaller fraction of 
hydrogenated carbons than in the MW (for a discussion, see \citealt{NersesianA&A2019}). This translates into an increase
in FIR emissivity due to the contribution of (larger) carbon grains. 
The position of the dots for each mock galaxy depends on the fitted value, and in part to a
(very small) effect of the temperature mixing. Nevertheless, within the uncertainties of the observations,
the mock results still follow those of the THEMIS model. We also tested cases with a larger variation in the heating conditions by removing
the $U_\mathrm{min}$ component and allowing all the dust to be heated by the $U^{-\alpha}$ distribution \citep[see e.g.][]{GallianoMNRAS2018}. Allowing $\alpha$ to vary between  1 and 3, we obtained fits to the DustPedia SEDs that are as good as those in \citet{NersesianA&A2019}. Their mock  $\emy$ versus $\emyrat$ trend (not shown) still follows the trend that was obtained from the original fits.
This shows that the mixing of heating conditions (dust temperatures) does not have a significant effect 
on the emissivity (and on the trend for the galaxies with $\emyrat > 6$). Unless the distribution 
of radiation fields we adopted is too simplistic and unable to describe the full range of heating 
conditions, the difference between the observed and modelled trends might indicate a difference
in the dust grain properties (possibly including a variation in the $D/H$ ratio) for galaxies of bluer FIR colours.

We recall that we did not include the contribution of \ion{H}{ii} to the gas column density.
For the diffuse MW medium, \ion{H}{ii} is estimated to account for $\approx$20\% of the total 
hydrogen content \citep{CompiegneA&A2011,DraineBook2011}. If this fraction does not
change between galaxies, the results presented here are not affected: 
 \ion{H}{ii} is neglected both in the DustPedia sample and in the MW reference emissivity.
Instead, if the fraction of ionised gas increases for galaxies with redder FIR colours, 
the $\emy$ trend across the whole $\emyrat$ range might be reconciled with the modelled trends.
Unfortunately, we were unable to find any systematic study on the 
variation of the fraction of \ion{H}{ii} in galaxies of different properties. When we use
the DL07 model as reference, passing through the average $\emy$ at higher $\emyrat$,
an increase of $\approx$40\% in $N_\mathrm{H}$ is needed at $\emyrat\approx4.5$ for the
sample average to fall on the trend (similar results can be found
using the other trends in the left panel of Fig.~\ref{fig:emy_vs_frat2}, rescaled at higher $\emyrat$
values). However, this
correction would be larger than what is estimated for the MW at the same $\emyrat$.
 $\emy$ is  even very weakly anti-correlated with galactic properties from \citet{NersesianA&A2019} that
trace the amount of ionising photons: it is only $\tau_\mathrm{K}=-0.11$ ($p_\mathrm{K}=3\%$)
for the correlation with the fraction of stellar luminosity coming from young stars (and similar
values for the $\mathit{sSFR}$ and the fraction of intrinsic unattenuated photons in the
FUV band). It is therefore unlikely that neglicting \ion{H}{ii} has significantly
affected the trend we presented here. The same conclusion might be applied to the contribution of
dark (molecular) gas, whose presence might not be revealed because the CO molecule in regions with stronger UV radiation fields is photodissociated
\citep{WolfireApJ2010}. Instead, unaccounted-for \ion{H}{ii} and dark gas in galaxies at higher 
$\emyrat$ would increase the deviation of the $\emy$ trend from the modelled ones 
more strongly than what we find here.

\begin{figure*}
\includegraphics[width=\hsize]{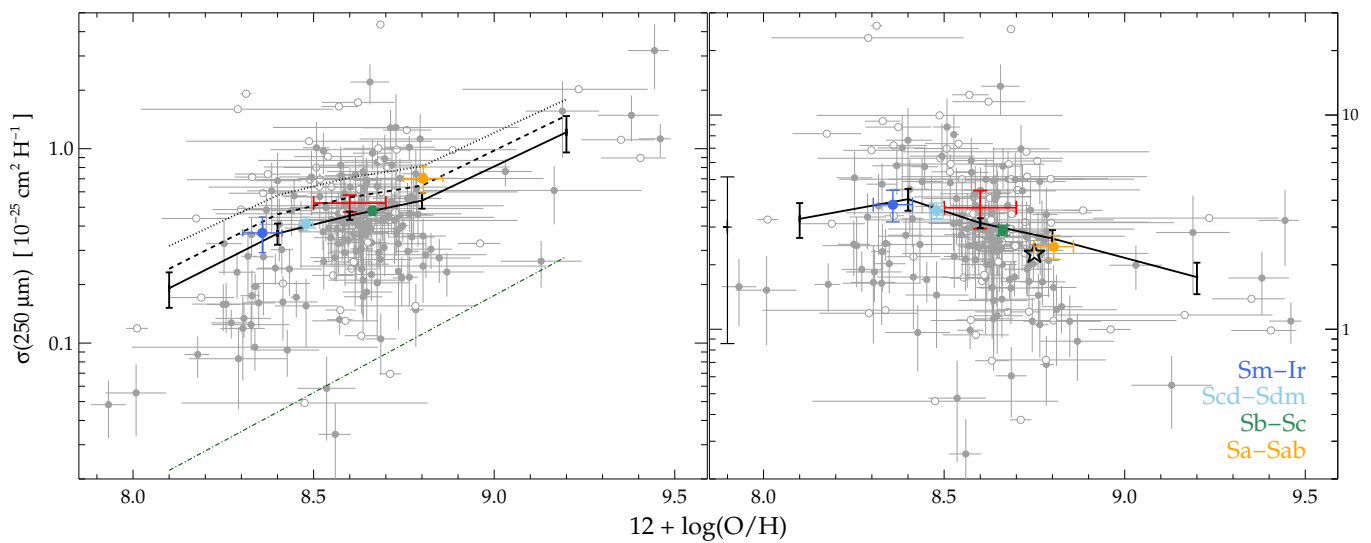}
\caption{
Absorption cross section per H atom $\absH$ (left panel) and per dust mass $\absM$ (right panel) as a function of the metallicity.
The two panels show individual measurements as grey dots, with the same convention for open and filled symbols as in Fig.~\ref{fig:emy_vs_frat1}.
The mean (and its standard deviation) of the quantity of the y-axis for the five metallicity bins defined in Fig.~\ref{fig:emy_vs_frat1a} (black error bars 
connected by a solid line) are shown, and the same, but for the four morphology bins (coloured dots with error bars). We also plot
the HeViCS MW estimate of Table~\ref{tab:mwrefs} (red error bars). In the left panel we also show the mean for each metallicity bin,
{
when the dust temperature is derived with
$\beta=1.8$ (dashed line) and 2.0 (dotted line)}, and 
the slope for a linear dependence of $\absM$ on $O/H$ (green dot-dashed line, arbitrary scale; Eqs.~\ref{eq:kappa} and~\ref{eq:mwscale}). In the right panel, 
the black error bar at the left shows the uncertainty in $((D/H)/(O/H))_\mathrm{MW}$. The star indicates the
value derived from C16 after rescaling it to the assumptions of this work (see text for details).
}
\label{fig:tau_vs_met}
\label{fig:k250_vs_met}
\end{figure*}

\section{Absorption cross section}
\label{sec:abs}

The derivation of the dust absorption cross-section is more uncertain because it requires assumptions 
on the heating field and on the $D/H$ ratio. We first discuss the absorption cross section per hydrogen 
atoms, $\absH$, where the effects of different dust  temperatures are removed. We then consider
the absorption cross section per dust mass, $\absM$, which further depends on the assumption 
on the $D/H$ ratio and its variations with metallicity. 

\subsection{Absorption cross section per hydrogen atom}

In Fig.~\ref{fig:tau_vs_met}  (left panel) we show $\absH$ as a function of the global gas metallicity. 
The dust temperature required to derive  $\absH$ (Eq.~\ref{eq:tau}) was obtained by fitting 
the SED with an MBB of $\beta=1.6$, the same value as derived from fits of the high galactic latitude cirrus. 
The black solid line and error bars show the  mean (and its standard deviation) for the same bins in metallicity 
as in Fig.~\ref{fig:emy_vs_frat1a}. Coloured dots with error bars show the same for the morphology bins. 
A mild trend of the absorption cross section with metallicity is visible that is compatible with the trend on 
morphology: objects of later type and lower metallicity have a smaller cross section than earlier type  higher metallicity 
spirals. The scatter is large, however, and the Kendall correlation coefficient is small ($\tau_\mathrm{K} = 0.21$, 
$p_\mathrm{K}=$0). This faint correlation might be partially 
{ due to the small number of  data points at low and high 
metallicity,  which have a larger CO-to-H$_2$ correction factor}: when it is restricted to the range $8.3 \le 12+\log(\mathrm{O/H}) < 8.9$,   
$\tau_\mathrm{K}$ reduces to 0.1 ($p_\mathrm{K}=$6\%). The correlation on this
restricted range disappears when a constant MW conversion factor is used
($\tau_\mathrm{K} = 0.02$, $p_\mathrm{K}=$75\%).

Even when we limit our analysis to the three central bins in metallicity, the average trend appears
flatter than (although marginally consistent with) the linear dependence of  $\absH$ on $O/H$
that is expected from the assumption of a universal dust-to-metal gas fraction 
(Eqs.~\ref{eq:kappa} and~\ref{eq:mwscale}; green dot-dashed line in Fig.~\ref{fig:tau_vs_met}, left panel).
However, we note that the trend could have been altered by choosing a single
$\beta$ for the whole sample. Using a larger $\beta$ 
would have resulted in lower $T_\mathrm{d}$ and higher
$\absH$: for $\beta=1.8$  (close to the value fitted to the THEMIS average cross section; 
\citealt{GallianoARA&A2018}), the estimate of $\absH$  rises by $\approx 20$\% (dashed line in 
Fig.~\ref{fig:tau_vs_met}, left panel); when $\beta=2$ (as in DL07; \citealt{BianchiA&A2013}), it is 55\% higher (dotted line). 
It has been shown that larger {\em \textup{apparent}} $\beta$ can describe the FIR to submm SED 
of galaxies with stronger radiation fields better, even without changes in the dust
composition \citep{HuntA&A2015}. An increase in  $\beta$ for earlier type galaxies with bluer
$\emyrat$ ratios could steepen the average trend of $\absH$ vs $O/H$  and
cause it to approach theoretical expectations. The observed trend might also be flatter 
because of the lower-than-expected $\emy$ of these galaxies, although the uncertainties 
associated with the temperature derivation do not make the effect as evident as in the
figures of Sect.~\ref{sec:emy}.
 
The red dot in Fig.~\ref{fig:tau_vs_met} (left panel) shows the value derived for the MW cirrus using HeViCS data. For the metallicity bin corresponding to the abundances we adopted for the MW gas (see Sect.~\ref{sec:mw}), the average value of $\absH$ of our sample is compatible with the MW estimate, but it is 15\% lower. The value becomes fully consistent with that of the
MW when the constant conversion factor of \citet{BolattoARA&A2013} is adopted or equivalently,
if the \citet{AmorinA&A2016} formula is rescaled to give the exact MW factor at our reference
metallicity  (see values in Table~\ref{tab:results}).
The value of  $\absH$ in the cirrus is also close to that of Sb-Sc galaxies (green dot in Fig.~\ref{fig:tau_vs_met}, left panel), a morphology range that is shared by the Galaxy 
(see Sect.~\ref{sec:emyres}). This is also in agreement with the estimates on the Galactic plane by the \citet{PlanckEarlyXXI}: after deriving the emissivity SED in
five Galactocentric rings and fitting them with an MBB, they found that $\absH$ at the solar circle is slightly lower than but compatible with  the value at high Galactic latitude.
Within the uncertainties in the fitting, they also detected no significant variation with Galactocentric radius and within the different ISM components. This confirms that
the dust properties in the cirrus are not significantly different from those in the disc.
 
For different MW regions, an anti-correlation between $\absH$ and $T_\mathrm{d}$ is found. This is generally interpreted as due to local variations in dust composition: for the same emitted 
power, an increase (decrease) in the absorption cross section results in a decrease (increase) of the 
temperature \citep{PlanckEarlyXXIV,MartinApJ2012,PlanckIntermediateXVII,Planck2013XI}. 
In optically opaque clouds, the use of a single dust temperature, neglecting the
temperature decrease from the outskirts to the core, might contribute to the effect
\citep{YsardA&A2012}.
An anti-correlation is also found in our galaxy sample. However, it is weak  ($\tau_\mathrm{K} = -0.17$, $p_\mathrm{K}=$0.03\%) and largely induced by the uncertainties in  $T_\mathrm{d}$.
Furthermore, in the MW the anti-correlation with $T_\mathrm{d}$ is also shared by the emissivity \citep{PlanckIntermediateXVII}, while in our sample the emissivity correlates positively with $T_\mathrm{d}$ ($\tau_\mathrm{K} = 0.20$, $p_\mathrm{K}=$0).

\subsection{Absorption cross section per dust mass}

The right panel of Fig.~\ref{fig:k250_vs_met} shows $\kappa (250\;\mu\mathrm{m})$ versus metallicity. The absorption cross section per dust mass was obtained from $\absH$ assuming a linear dependence of the $D/H$ ratio on metallicity (Eq.~\ref{eq:mwscale}). All errors were propagated to  $\kappa (250\;\mu\mathrm{m})$, except for the uncertainty in the $((D/H)/(O/H))_\mathrm{MW}$ normalisation, which applies to all data points (it is shown as the black error bar at the left of  Fig.~\ref{fig:k250_vs_met}, right panel). As for the other panel, we plot
the average for metallicity bins, as well as that for bins in morphology (which share the same trend as those in metallicity, as we saw before). 
Like $\absH$, the value of $\absM$ for the central bin in metallicity is lower than but consistent with the MW estimate from the HeViCS cirrus (and becomes closer to it for a constant CO-to-H$_2$ conversion factor; see the values in Table~\ref{tab:results}).

As we discussed above, our method for deriving $\absM$ is analogous to that of C16. The main differences are in their choice of reference wavelength (500 $\mu$m), CO-to-H$_2$ conversion (from \citealt{SchrubaAJ2012}, with O3N2 metallicity, calibrated as in \citealt{PettiniMNRAS2004}), dust emission model (two-T MBB with $\beta=2$, although the cold component is dominant), and assumption for the MW value for the $(D/H)/(O/H)$ ratio (see the appendix). When we use the DustPedia dataset (that also includes O3N2 metallicities) and the  C16 recipe for the 15 galaxies in common with DustPedia (out of their total sample of 22), we obtain a mean value that agrees with theirs within 10\% (although results for individual galaxies differ more because of differences in the data). The galaxies in C16 have metallicities $8.6 < 12+\log_{10} (O/H) < 8.9$. Selecting DustPedia objects in the same range, we derived an average correction factor between our values and those obtained assuming the same recipe as  C16 (but with a single $T_\mathrm{d}$). With this factor,  the C16 absorption cross section becomes $\kappa (250\;\mu\mathrm{m}) = 2.3 \pm 0.1 \; \mathrm{cm^2 \; g^{-1}}$ (black star in Fig.~\ref{fig:k250_vs_met}, right panel). The value is lower than but still marginally consistent with our estimates.

The trend of $\absM$  is obviously degenerate with our assumption of a linear dependence 
of $D/H$ on metallicity: the (marginal) positive correlation of $\absH$ with metallicity becomes negative for $\absM$, with $\tau_\mathrm{H}$ = -0.19 ($p_\mathrm{K}=$0.01\%). However, 
the correlation might be induced by the intrinsic dependence of $\absM$ on (the inverse of) $O/H$ 
and the large uncertainties in metallicity. We verified the trend using proxies of metallicity that were not directly
related to the observables we used to derive $\absM$: for our sample, $M_\star$ and $1/\mathit{sSFR}$
are the quantities that correlate best with metallicity ($\tau_\mathrm{K} =0.43$ and 0.31, respectively, with
$p_\mathrm{K}=$0 in both cases); this is expected because of the mass-metallicity \citep{TremontiApJ2004}
and fundamental metallicity \citep{MannucciMNRAS2010} relations. For $\absM$ vs $M_\star$, the
strength of the trend is reduced ($\tau_\mathrm{K}$=0.14, $p_\mathrm{K}=$0.2\%), while no significant correlation 
is found with  $1/\mathit{sSFR}$ ($\tau_\mathrm{K}$=-0.01, with a probability for the null hypothesis $p_\mathrm{K}=$77\%).
The results are qualitatively similar when we change our reference CO-to-H$_2$ correction factor to  a  
constant one, or when we use a stronger quadratic dependence of the conversion factor on metallicity (such as that derived 
by \citealt{HuntA&A2015b}). 
The results shown in Fig.~\ref{fig:k250_vs_met} (right panel) might thus be consistent with no variation 
of $\absM$ with metallicity, or equivalently,   with the assumption that the fraction of metals 
included in grains is universal. 

Instead, theoretical models suggest that the dust-to-metal ratio takes the imprint of the dominant dust-formation mechanism and results in a different dependence of $D/H$ on metallicity \citep{MattssonMNRAS2012,AsanoEP&S2013}. \citet{RemyRuyerA&A2014} found that below a critical metallicity ($12 + \log(\mathrm{O/H}) \la 8$), the dependence of $D/H$ is steeper than linear. A similar conclusion was drawn  by \citet{VilchezMNRAS2019}, who analysed resolved observations of NGC628 and M101. However, these metallicities are absent (or underrepresented for the critical metallicity of 8.4 in \citealt{VilchezMNRAS2019}) from our sample. Nevertheless, the scatter is large and other works have found  $D/H \sim (O/H)^{1.5-2}$ over the full metallicity range in resolved studies (in the MW, \citealt{GiannettiA&A2017}; and M101, \citealt{ChiangApJ2018}) and global studies \citep[][although at higher metallicity their results are consistent with a constant dust-to-metal ratio]{DeVisA&A2019}. 
Adopting a steeper dependence of $D/H$ on metallicity would cause $\absM$ to be even more negatively correlated with $O/H$. However, these results depend on various assumptions and ultimately on the
use of the same $\absM$ for all the objects and environments. Paradoxically, they would require a positive correlation of $\absH$ on $O/H$ steeper than what is actually shown in the left panel 
of Fig.~\ref{fig:tau_vs_met}.

\subsection{Comparison with dust grain models}

\begin{figure}
\includegraphics[width=\hsize]{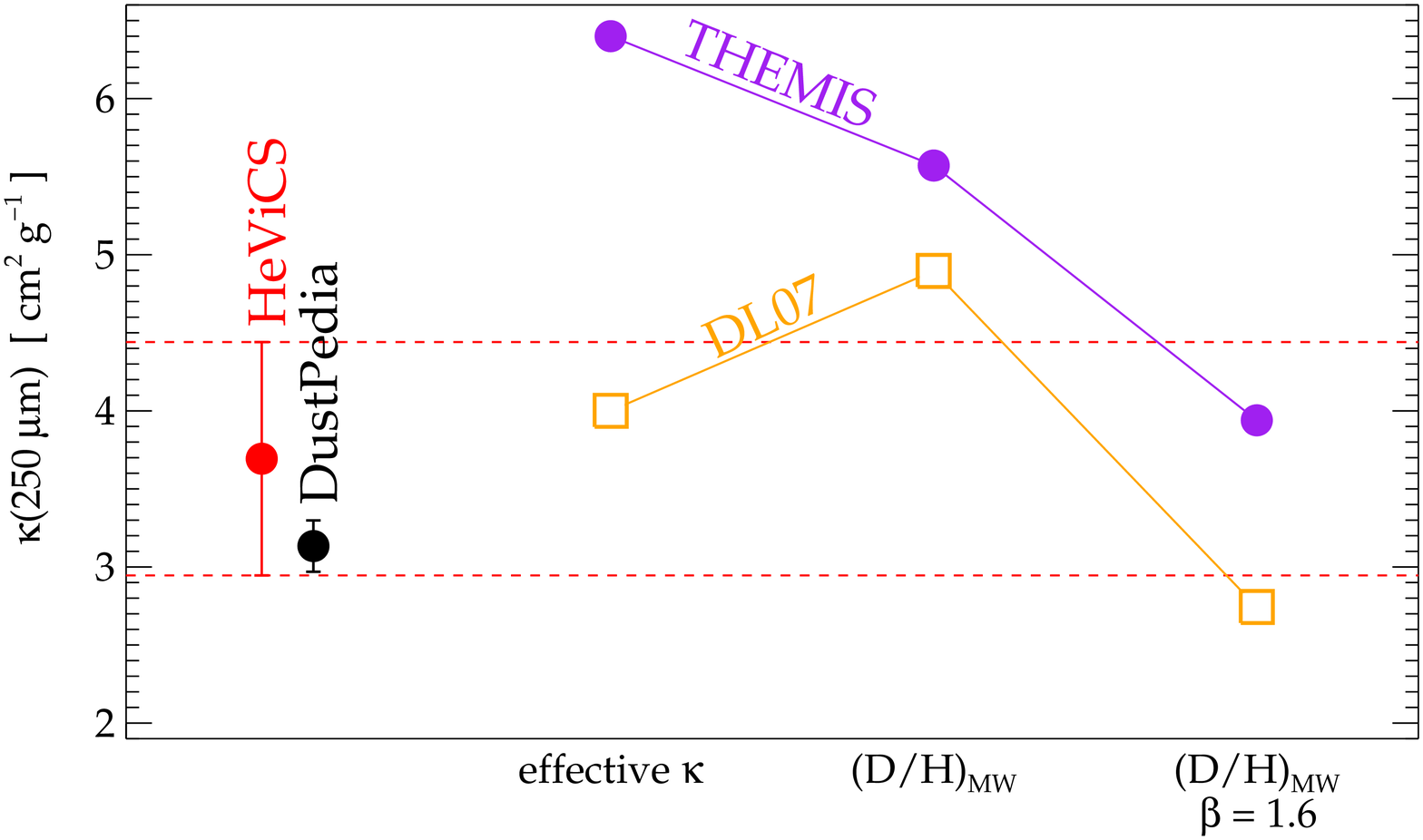}
\caption{
Absorption cross section per dust mass $\absM$ for the MW (red error bar) and for our sample in the MW metallicity bin (black error bar) compared with
the values expected from the THEMIS and DL07 grain models (under various assumptions for the $D/H$ ratio and $\beta$; see text for details).
}
\label{fig:k250_vs_models}
\end{figure}

As a result of the fit to various constraints, dust models are characterised by an {\em \textup{effective}} 
$\kappa_\nu$, that is, by the average of the absorption cross section over the grain size distribution 
and material mix. Typically, the effective $\kappa_\nu$ in the FIR can be described by a power law such
as that of Eq.~\ref{eq:tau_beta} (see Table~\ref{tab:mwrefs}). Moreover, for the chosen grain materials, 
each model predicts its own $D/H$ ratio, constrained by the observed metal depletions. When 
we compare our results for $\absM$ with those of the models, we need to take into account these
differences in $\beta$ and $D/H$. We did this in Fig.~\ref{fig:k250_vs_models}, where we show
$\absM$ for the MW and for our sample at the MW metallicity, together with the predictions from 
THEMIS and DL07. 

Despite the common value for $\emy$, the THEMIS model (purple circles in Fig.~\ref{fig:k250_vs_models})
has a higher effective $\absM$ than our 
measurements. When we scale the model from its original value for $D/H$ to the adopted $(D/H)_\mathrm{MW}$ (which is equivalent to comparing their respective $\absH$), the difference
is smaller. The rescaled THEMIS is still about 2$\sigma$ away from the MW value. This gap
can be further reduced by considering a common $\beta$ for observations and model. The MW value for $\absM$ was derived after fitting the local emissivity SED \citep{BianchiA&A2017}. 
For a proper comparison, the emissivity predicted by the model when heated by the LISRF ($U=1$) 
should be fitted accordingly. Using an MBB and allowing the spectral index to vary as well, 
 \citet{KoehlerA&A2015} found $\beta \approx 1.5$. The {\em \textup{apparent}} $\beta$ is
smaller than that which describes the effective absorption cross section of the model, $\beta\approx 1.8$.
The difference is due to temperature mixing: the SED of the model results from the
composition of MBBs at the temperatures attained by grains of different materials and sizes,
while the fit considers only one temperature and tries to match the flatter slope by varying both temperature and $\beta$.
The effect is stronger (and the fitted apparent  $\beta$ smaller than the effective) for 
radiation fields with a lower intensity \citep{HuntA&A2015}. Using $\beta=1.6$, as for the MW and
DustPedia estimates, we fitted the $U=1$ THEMIS emissivity for 100 < $\lambda / \mu$m < 500
and further reduced the estimate for the model $\absM$. When both the $O/H$ and $\beta$ are
matched, the $\absM$ fitted to the model is  very close to the Galactic value (and to the 
DustPedia sample average, after the discussion in the previous sections is taken into account). 
This is expected because the emissivity SED predicted by THEMIS for the MW is not far different from
the observed one (Fig.~\ref{fig:emysed}).

\begin{table*}
\caption{Emissivity and absorption cross sections for our sample. Values in bold are those
given in the figures.}              
\label{tab:results}      
\small
\centering                                      
\begin{tabular}{c c l  l }          
\hline\hline                        
Quantity & \multicolumn{2}{c}{value}  & notes  \\
\hline                                   
\rule{0pt}{3ex} 
$\emy$ & $\mathbf{ 0.82 \pm 0.07}$ & MJy sr$^{-1}$ ($10^{20}$ H cm$^{-2}$)$^{-1}$ &   $\emyrat\approx4.5$   \\
             & $0.83 \pm 0.07$ &  &  (CO-to-H$_2$)$_\mathrm{MW}$  \\
             & $\mathbf{ 1.7   \pm 0.2}$&  &  $\emyrat\approx7.5$  \\
\rule{0pt}{3ex} 
$\absH$    &  $\mathbf{ 0.45 \pm 0.02}$ & $10^{-25}$ cm$^2$ H$^{-1}$ &  $(O/H)_\mathrm{MW}$ \\
                 & $0.52 \pm 0.02$                  &       & (CO-to-H$_2$)$_\mathrm{MW}$  \\
\rule{0pt}{3ex} 
$\absM$    & $\mathbf{ 3.1 \pm 0.2}$ & $\mathrm{cm}^2 \;\mathrm{g}^{-1}$ & $(O/H)_\mathrm{MW}$ \\
                  & $3.5 \pm 0.2$ &  & (CO-to-H$_2$)$_\mathrm{MW}$     \rule[-1.2ex]{0pt}{0pt}\\
\hline                                             
\end{tabular}
\end{table*}

\citet{NersesianA&A2019} have shown that the dust masses of DustPedia galaxies can be
retrieved equally well (within the observational uncertainties) using the model 
effective $\absM$ coupled with the MBB approximation, or the full model approach accounting
for distribution in grain sizes, materials, and ISRF intensities (see also \citealt{BianchiA&A2013}).
We tested what happens when we use the value of $\absM$ obtained from the fit above rescaled
to $(D/H)_\mathrm{MW}$. In this case the dust masses for DustPedia galaxies are $\approx 30\%$
higher 
than when we use the unscaled THEMIS absorption cross section.
Half of the relative difference 
is due to the different choice for $D/H$ and half to the 
fixed-$\beta$ MBB approximations. We repeated similar tests for the DL07 model (orange squares 
in Fig.~\ref{fig:k250_vs_models}). This time, the adoption of $(D/H)_\mathrm{MW}$ (lower
that the model value) results in a higher $\absM$ and dust masses lower by $\approx 18\%$.
The use of the fixed-$\beta$ MBB approximation instead leads to dust masses higher by
$\approx 12\%$: the net effect of both choices is a moderate underestimate of the dust mass
by $\approx 6\%$ 
with respect to the original DL07 model.
The lower predicted  emissivity of DL07  (Fig.~\ref{fig:emysed}) results in higher dust mass
estimates than were obtained with THEMIS. Even when $D/H$ 
and $\beta$ are matched, dust masses obtained from the DL07 model would be $\approx 45\%$ 
higher than those from THEMIS (because of a correspondingly lower $\absM$). When the original 
models are used, the DL07-derived masses are about a factor of 3 higher than for THEMIS,
as shown by  \citet{NersesianA&A2019} for the DustPedia sample.
The low DL07 emissivity is likely the cause of the overestimate of the optical/near-infrared
extinction when compared to direct extinction measurements
in the MW  \citep{PlanckIntermediateXXIX} and the Andromeda Galaxy \citep{DalcantonApJ2015}.

The comparison we made with models shows the potential pitfalls of deriving the absorption 
cross section from FIR observations, which is in particular due to the uncertainties on the $D/H$ ratio 
and on the MBB approximation. This strengthens our belief that a more reliable characterisation
of the dust properties in external galaxies should be made at the $\emy$ level. This quantity
relies on fewer assumptions. This is after all what is commonly done 
to constrain models for dust in the MW cirrus.

\section{Summary and conclusions}
\label{sec:summa}

We estimated the average dust emissivity $\emy$ for 204 late-type galaxies belonging to the DustPedia sample \citep{DaviesPASP2017}. All our objects have multi-wavelength photometry \citep{ClarkA&A2018}, SED-fitted models, and estimates for physical quantities \citep{NersesianA&A2019}, metallicity \citep{DeVisA&A2019}, and atomic and molecular gas data (\citealt{DeVisA&A2019}; Casasola et al., submitted). Using the single-T MBB approximation, we derived the absorption cross sections per H-atom $\absH$; and per mass $\absM$ by adopting a dust-to-hydrogen mass ratio $D/H$. The results are summarised in Table~\ref{tab:results}, and the assumptions and MW reference 
values are listed in Table~\ref{tab:mwrefs}. Our main findings are listed below.

\begin{enumerate}

\item In galaxies with global FIR colours similar to that of the Galactic cirrus, $\emy$ is very close to that measured in the MW and consistent with predictions from the THEMIS model (and from an MBB  with
$\beta=1.6$).  The result 
is almost independent of the main assumption in the emissivity derivation, that is,\ the variation of the CO-to-H$_2$ conversion factor with metallicity. This reassures us that MW-based dust models can be used to derive the mass in external galaxies.

\item For galaxies with bluer FIR colours, $\emy$ is on average smaller than what is implied by its dependence on temperature. These objects are typically  of earlier type and higher metallicity,
and have an H$_2$  component that
contributes significantly if not dominantly 
to the global gas mass. 
This might suggest possible variations in dust properties with the environment or the lack (of detection) of dust associated with H$_2$.

\item For objects in the metallicity range of MW gas, the average absorption cross sections, either per  H atom, $\absH$ or per dust mass, $\absM$, are slightly smaller than (by 15\%) but compatible with cirrus estimates.
The results would become entirely consistent with MW values for a constant  CO-to-H$_2$ conversion factor. Mild trends are seen with $O/H$, although the observational scatter and the uncertainties in the assumptions
prevent a  definitive assessment of the variation of the cross sections with metallicity.

\end{enumerate}

One of the main assumptions in this work, as well as in almost all determinations of the dust mass in galaxies, is that the same dust emission properties are valid throughout the whole disc. This appears to be the case for the MW, where  $\absH$ is found to be the same within each of the gas components on the Galactic plane, and at high latitudes \citep{PlanckEarlyXXI}. In this work, we instead found indications that the dust emissivity for galaxies is reduced whose gas mass is dominated by H$_2$. A more definitive
assessment of the variation of dust properties with the environment requires resolved observations of dust, gas, and metallicity. In parallel to the current work, a pilot study has been conducted on two DustPedia galaxies, M74 and M83: reassuringly, the dust absorption cross section is reduced in these objects in regions of higher gas column density. This qualitatively agrees with our findings \citep{ClarkMNRAS2019}.

Taken at face value, the lower values of $\emy$ might indicate without accounting for any $D/H$ variation that the dust masses for galaxies with bluer FIR colours are underestimated by up to a factor two.
While we wait for other observational confirmations of this result, it might be interesting to understand from the theoretical viewpoint whether the exposition of dust grains to strong radiation fields can cause these
variations \citep{JonesA&A2017}. Before any other modelling is undertaken, however, we recommend a revision and standardisation of the observational constraints. Comparing the predictions of two 
commonly used dust grain models, we find discrepancies in the adopted cirrus emissivity, related to both its absolute value and the correction for the \ion{H}{ii} contribution (accounting for this could reduce $\emy$ by 20\%).
Other uncertainties might lie in the definition of the LISFR: \citet{DraineBook2011} proposed a revisited version of the commonly used spectrum from \citet{MathisA&A1983}, with an energy output increased by about 30\% 
\citep{BianchiA&A2017}. Finally, a revision is recommended also for the optical extinction per H column density, which is required to scale another main constraint for dust models, the extinction law:
the commonly used value \citep{BohlinApJ1978} is 40 - 60\% higher than the newer estimates  \citep{LisztApJ2014,LenzApJ2017,NguyenApJ2018}
 
\begin{acknowledgements}
We would like to thank the anonymous referee and G. Cresci for useful comments, and L.~Verstraete for help with the \citet{CompiegneA&A2011} emissivity. S.~B., V.~C., L.~M., and E.~C.  acknowledge funding from the INAF mainstream 2018 program "Gas-DustPedia: A definitive view of the ISM in the Local Universe".
\end{acknowledgements}

\bibliographystyle{aa} 
\bibliography{/Users/sbianchi/Documents/tex/DUST} 

\begin{thebibliography}{89}
\expandafter\ifx\csname natexlab\endcsname\relax\def\natexlab#1{#1}\fi

\bibitem[{{Amor{\'{\i}}n} {et~al.}(2016){Amor{\'{\i}}n},
  {Mu{\~n}oz-Tu{\~n}{\'o}n}, {Aguerri}, \& {Planesas}}]{AmorinA&A2016}
{Amor{\'{\i}}n}, R., {Mu{\~n}oz-Tu{\~n}{\'o}n}, C., {Aguerri}, J.~A.~L., \&
  {Planesas}, P. 2016, A\&A, 588, A23

\bibitem[{{Arendt} {et~al.}(1998){Arendt}, {Odegard}, {Weiland}, {Sodroski},
  {Hauser}, {Dwek}, {Kelsall}, {Moseley}, {Silverberg}, {Leisawitz},
  {Mitchell}, {Reach}, \& {Wright}}]{ArendtApJ1998}
{Arendt}, R.~G., {Odegard}, N., {Weiland}, J.~L., {et~al.} 1998, ApJ, 508, 74

\bibitem[{{Asano} {et~al.}(2013){Asano}, {Takeuchi}, {Hirashita}, \&
  {Inoue}}]{AsanoEP&S2013}
{Asano}, R.~S., {Takeuchi}, T.~T., {Hirashita}, H., \& {Inoue}, A.~K. 2013,
  Earth, Planets, and Space, 65, 213

\bibitem[{{Bianchi}(2013)}]{BianchiA&A2013}
{Bianchi}, S. 2013, A\&A, 552, A89

\bibitem[{{Bianchi} {et~al.}(2018){Bianchi}, {De Vis}, {Viaene}, {Nersesian},
  {Mosenkov}, {Xilouris}, {Baes}, {Casasola}, {Cassar{\`a}}, {Clark}, {Davies},
  {De Looze}, {Dobbels}, {Galametz}, {Galliano}, {Jones}, {Lianou}, {Madden},
  \& {Tr{\v c}ka}}]{BianchiA&A2018}
{Bianchi}, S., {De Vis}, P., {Viaene}, S., {et~al.} 2018, A\&A, 620, A112

\bibitem[{{Bianchi} {et~al.}(2017){Bianchi}, {Giovanardi}, {Smith}, {Fritz},
  {Davies}, {Haynes}, {Giovanelli}, {Baes}, {Bocchio}, {Boissier}, {Boquien},
  {Boselli}, {Casasola}, {Clark}, {De Looze}, {di Serego Alighieri}, {Grossi},
  {Jones}, {Hughes}, {Hunt}, {Madden}, {Magrini}, {Pappalardo}, {Ysard}, \&
  {Zibetti}}]{BianchiA&A2017}
{Bianchi}, S., {Giovanardi}, C., {Smith}, M.~W.~L., {et~al.} 2017, A\&A, 597,
  A130

\bibitem[{{Bocchio} {et~al.}(2014){Bocchio}, {Jones}, \&
  {Slavin}}]{BocchioA&A2014}
{Bocchio}, M., {Jones}, A.~P., \& {Slavin}, J.~D. 2014, A\&A, 570, A32

\bibitem[{{Boggess} {et~al.}(1992){Boggess}, {Mather}, {Weiss}, {Bennett},
  {Cheng}, {Dwek}, {Gulkis}, {Hauser}, {Janssen}, {Kelsall}, {Meyer},
  {Moseley}, {Murdock}, {Shafer}, {Silverberg}, {Smoot}, {Wilkinson}, \&
  {Wright}}]{BoggessApJ1992}
{Boggess}, N.~W., {Mather}, J.~C., {Weiss}, R., {et~al.} 1992, ApJ, 397, 420

\bibitem[{{Bohlin} {et~al.}(1978){Bohlin}, {Savage}, \&
  {Drake}}]{BohlinApJ1978}
{Bohlin}, R.~C., {Savage}, B.~D., \& {Drake}, J.~F. 1978, ApJ, 224, 132

\bibitem[{{Bolatto} {et~al.}(2013){Bolatto}, {Wolfire}, \&
  {Leroy}}]{BolattoARA&A2013}
{Bolatto}, A.~D., {Wolfire}, M., \& {Leroy}, A.~K. 2013, ARA\&A, 51, 207

\bibitem[{{Boquien} {et~al.}(2019){Boquien}, {Burgarella}, {Roehlly}, {Buat},
  {Ciesla}, {Corre}, {Inoue}, \& {Salas}}]{BoquienA&A2019}
{Boquien}, M., {Burgarella}, D., {Roehlly}, Y., {et~al.} 2019, \aap, 622, A103

\bibitem[{{Casasola} {et~al.}(2017){Casasola}, {Cassar{\`a}}, {Bianchi},
  {Verstocken}, {Xilouris}, {Magrini}, {Smith}, {De Looze}, {Galametz},
  {Madden}, {Baes}, {Clark}, {Davies}, {De Vis}, {Evans}, {Fritz}, {Galliano},
  {Jones}, {Mosenkov}, {Viaene}, \& {Ysard}}]{CasasolaA&A2017}
{Casasola}, V., {Cassar{\`a}}, L.~P., {Bianchi}, S., {et~al.} 2017, A\&A, 605,
  A18

\bibitem[{{Chastenet} {et~al.}(2017){Chastenet}, {Bot}, {Gordon}, {Bocchio},
  {Roman-Duval}, {Jones}, \& {Ysard}}]{ChastenetA&A2017}
{Chastenet}, J., {Bot}, C., {Gordon}, K.~D., {et~al.} 2017, A\&A, 601, A55

\bibitem[{{Chiang} {et~al.}(2018){Chiang}, {Sandstrom}, {Chastenet}, {Johnson},
  {Leroy}, \& {Utomo}}]{ChiangApJ2018}
{Chiang}, I.-D., {Sandstrom}, K.~M., {Chastenet}, J., {et~al.} 2018, ApJ, 865,
  117

\bibitem[{{Clark} {et~al.}(2019){Clark}, {De Vis}, {Baes}, \&
  {DustPedia}}]{ClarkMNRAS2019}
{Clark}, C.~J.~R., {De Vis}, P., {Baes}, M., \& {DustPedia}, A. 2019, MNRAS,
  accepted

\bibitem[{{Clark} {et~al.}(2016){Clark}, {Schofield}, {Gomez}, \&
  {Davies}}]{ClarkMNRAS2016}
{Clark}, C.~J.~R., {Schofield}, S.~P., {Gomez}, H.~L., \& {Davies}, J.~I. 2016,
  MNRAS, 459, 1646

\bibitem[{{Clark} {et~al.}(2018){Clark}, {Verstocken}, {Bianchi}, {Fritz},
  {Viaene}, {Smith}, {Baes}, {Casasola}, {Cassara}, {Davies}, {De Looze}, {De
  Vis}, {Evans}, {Galametz}, {Jones}, {Lianou}, {Madden}, {Mosenkov}, \&
  {Xilouris}}]{ClarkA&A2018}
{Clark}, C.~J.~R., {Verstocken}, S., {Bianchi}, S., {et~al.} 2018, A\&A, 609,
  A37

\bibitem[{{Compi{\`e}gne} {et~al.}(2011){Compi{\`e}gne}, {Verstraete}, {Jones},
  {Bernard}, {Boulanger}, {Flagey}, {Le Bourlot}, {Paradis}, \&
  {Ysard}}]{CompiegneA&A2011}
{Compi{\`e}gne}, M., {Verstraete}, L., {Jones}, A., {et~al.} 2011, A\&A, 525,
  A103

\bibitem[{{Curti} {et~al.}(2017){Curti}, {Cresci}, {Mannucci}, {Marconi},
  {Maiolino}, \& {Esposito}}]{CurtiMNRAS2017}
{Curti}, M., {Cresci}, G., {Mannucci}, F., {et~al.} 2017, MNRAS, 465, 1384

\bibitem[{{Dalcanton} {et~al.}(2015){Dalcanton}, {Fouesneau}, {Hogg}, {Lang},
  {Leroy}, {Gordon}, {Sand strom}, {Weisz}, {Williams}, {Bell}, {Dong},
  {Gilbert}, {Gouliermis}, {Guhathakurta}, {Lauer}, {Schruba}, {Seth}, \&
  {Skillman}}]{DalcantonApJ2015}
{Dalcanton}, J.~J., {Fouesneau}, M., {Hogg}, D.~W., {et~al.} 2015, ApJ, 814, 3

\bibitem[{{Dale} {et~al.}(2012){Dale}, {Aniano}, {Engelbracht}, {Hinz},
  {Krause}, {Montiel}, {Roussel}, {Appleton}, {Armus}, {Beir{\~a}o}, {Bolatto},
  {Brandl}, {Calzetti}, {Crocker}, {Croxall}, {Draine}, {Galametz}, {Gordon},
  {Groves}, {Hao}, {Helou}, {Hunt}, {Johnson}, {Kennicutt}, {Koda}, {Leroy},
  {Li}, {Meidt}, {Miller}, {Murphy}, {Rahman}, {Rix}, {Sandstrom}, {Sauvage},
  {Schinnerer}, {Skibba}, {Smith}, {Tabatabaei}, {Walter}, {Wilson}, {Wolfire},
  \& {Zibetti}}]{DaleApJ2012}
{Dale}, D.~A., {Aniano}, G., {Engelbracht}, C.~W., {et~al.} 2012, ApJ, 745, 95

\bibitem[{{Davies} {et~al.}(2010){Davies}, {Baes}, {Bendo}, {Bianchi},
  {Bomans}, {Boselli}, {Clemens}, {Corbelli}, {Cortese}, {Dariush}, {de Looze},
  {di Serego Alighieri}, {Fadda}, {Fritz}, {Garcia-Appadoo}, {Gavazzi},
  {Giovanardi}, {Grossi}, {Hughes}, {Hunt}, {Jones}, {Madden}, {Pierini},
  {Pohlen}, {Sabatini}, {Smith}, {Verstappen}, {Vlahakis}, {Xilouris}, \&
  {Zibetti}}]{DaviesA&A2010}
{Davies}, J.~I., {Baes}, M., {Bendo}, G.~J., {et~al.} 2010, A\&A, 518, L48

\bibitem[{{Davies} {et~al.}(2017){Davies}, {Baes}, {Bianchi}, {Jones},
  {Madden}, {Xilouris}, {Bocchio}, {Casasola}, {Cassara}, {Clark}, {De Looze},
  {Evans}, {Fritz}, {Galametz}, {Galliano}, {Lianou}, {Mosenkov}, {Smith},
  {Verstocken}, {Viaene}, {Vika}, {Wagle}, \& {Ysard}}]{DaviesPASP2017}
{Davies}, J.~I., {Baes}, M., {Bianchi}, S., {et~al.} 2017, PASP, 129, 044102

\bibitem[{{De Vis} {et~al.}(2019){De Vis}, {Jones}, {Viaene}, {Casasola},
  {Clark}, {Baes}, {Bianchi}, {Cassara}, {Davies}, {De Looze}, {Galametz},
  {Galliano}, {Lianou}, {Madden}, {Manilla-Robles}, {Mosenkov}, {Nersesian},
  {Roychowdhury}, {Xilouris}, \& {Ysard}}]{DeVisA&A2019}
{De Vis}, P., {Jones}, A., {Viaene}, S., {et~al.} 2019, A\&A, 623, A5

\bibitem[{{D\'esert} {et~al.}(1990){D\'esert}, {Boulanger}, \&
  {Puget}}]{DesertA&A1990}
{D\'esert}, F.~X., {Boulanger}, F., \& {Puget}, J.~L. 1990, A\&A, 237, 215

\bibitem[{{Draine}(2003)}]{DraineARA&A2003}
{Draine}, B.~T. 2003, ARA\&A, 41, 241

\bibitem[{{Draine}(2011)}]{DraineBook2011}
{Draine}, B.~T. 2011, {Physics of the Interstellar and Intergalactic Medium}
  ({Princeton University Press})

\bibitem[{{Draine} {et~al.}(2007){Draine}, {Dale}, {Bendo}, {Gordon}, {Smith},
  {Armus}, {Engelbracht}, {Helou}, {Kennicutt}, {Li}, {Roussel}, {Walter},
  {Calzetti}, {Moustakas}, {Murphy}, {Rieke}, {Bot}, {Hollenbach}, {Sheth}, \&
  {Teplitz}}]{DraineApJ2007}
{Draine}, B.~T., {Dale}, D.~A., {Bendo}, G., {et~al.} 2007, ApJ, 663, 866

\bibitem[{{Draine} \& {Lee}(1984)}]{DraineApJ1984}
{Draine}, B.~T. \& {Lee}, H.~M. 1984, ApJ, 285, 89

\bibitem[{{Draine} \& {Li}(2007)}]{DraineApJ2007b}
{Draine}, B.~T. \& {Li}, A. 2007, ApJ, 657, 810

\bibitem[{{Dwek} {et~al.}(1997){Dwek}, {Arendt}, {Fixsen}, {Sodroski},
  {Odegard}, {Weiland}, {Reach}, {Hauser}, {Kelsall}, {Moseley}, {Silverberg},
  {Shafer}, {Ballester}, {Bazell}, \& {Isaacman}}]{DwekApJ1997}
{Dwek}, E., {Arendt}, R.~G., {Fixsen}, D.~J., {et~al.} 1997, ApJ, 475, 565

\bibitem[{{Esteban} \& {Garc{\'{\i}}a-Rojas}(2018)}]{EstebanMNRAS2018}
{Esteban}, C. \& {Garc{\'{\i}}a-Rojas}, J. 2018, MNRAS, 478, 2315

\bibitem[{{Galliano}(2018)}]{GallianoMNRAS2018}
{Galliano}, F. 2018, MNRAS, 476, 1445

\bibitem[{{Galliano} {et~al.}(2018){Galliano}, {Galametz}, \&
  {Jones}}]{GallianoARA&A2018}
{Galliano}, F., {Galametz}, M., \& {Jones}, A.~P. 2018, ARA\&A, 56, 673

\bibitem[{{Giannetti} {et~al.}(2017){Giannetti}, {Leurini}, {K{\"o}nig},
  {Urquhart}, {Pillai}, {Brand}, {Kauffmann}, {Wyrowski}, \&
  {Menten}}]{GiannettiA&A2017}
{Giannetti}, A., {Leurini}, S., {K{\"o}nig}, C., {et~al.} 2017, A\&A, 606, L12

\bibitem[{{Griffin} {et~al.}(2010){Griffin}, {Abergel}, {Abreu}, {Ade},
  {Andr{\'e}}, {Augueres}, {Babbedge}, {Bae}, {Baillie}, {Baluteau}, {Barlow},
  {Bendo}, {Benielli}, {Bock}, {Bonhomme}, {Brisbin}, {Brockley-Blatt},
  {Caldwell}, {Cara}, {Castro-Rodriguez}, {Cerulli}, {Chanial}, {Chen},
  {Clark}, {Clements}, {Clerc}, {Coker}, {Communal}, {Conversi}, {Cox},
  {Crumb}, {Cunningham}, {Daly}, {Davis}, {de Antoni}, {Delderfield}, {Devin},
  {di Giorgio}, {Didschuns}, {Dohlen}, {Donati}, {Dowell}, {Dowell}, {Duband},
  {Dumaye}, {Emery}, {Ferlet}, {Ferrand}, {Fontignie}, {Fox}, {Franceschini},
  {Frerking}, {Fulton}, {Garcia}, {Gastaud}, {Gear}, {Glenn}, {Goizel},
  {Griffin}, {Grundy}, {Guest}, {Guillemet}, {Hargrave}, {Harwit}, {Hastings},
  {Hatziminaoglou}, {Herman}, {Hinde}, {Hristov}, {Huang}, {Imhof}, {Isaak},
  {Israelsson}, {Ivison}, {Jennings}, {Kiernan}, {King}, {Lange}, {Latter},
  {Laurent}, {Laurent}, {Leeks}, {Lellouch}, {Levenson}, {Li}, {Li},
  {Lilienthal}, {Lim}, {Liu}, {Lu}, {Madden}, {Mainetti}, {Marliani}, {McKay},
  {Mercier}, {Molinari}, {Morris}, {Moseley}, {Mulder}, {Mur}, {Naylor},
  {Nguyen}, {O'Halloran}, {Oliver}, {Olofsson}, {Olofsson}, {Orfei}, {Page},
  {Pain}, {Panuzzo}, {Papageorgiou}, {Parks}, {Parr-Burman}, {Pearce},
  {Pearson}, {P{\'e}rez-Fournon}, {Pinsard}, {Pisano}, {Podosek}, {Pohlen},
  {Polehampton}, {Pouliquen}, {Rigopoulou}, {Rizzo}, {Roseboom}, {Roussel},
  {Rowan-Robinson}, {Rownd}, {Saraceno}, {Sauvage}, {Savage}, {Savini},
  {Sawyer}, {Scharmberg}, {Schmitt}, {Schneider}, {Schulz}, {Schwartz},
  {Shafer}, {Shupe}, {Sibthorpe}, {Sidher}, {Smith}, {Smith}, {Smith},
  {Spencer}, {Stobie}, {Sudiwala}, {Sukhatme}, {Surace}, {Stevens}, {Swinyard},
  {Trichas}, {Tourette}, {Triou}, {Tseng}, {Tucker}, {Turner}, {Vaccari},
  {Valtchanov}, {Vigroux}, {Virique}, {Voellmer}, {Walker}, {Ward}, {Waskett},
  {Weilert}, {Wesson}, {White}, {Whitehouse}, {Wilson}, {Winter}, {Woodcraft},
  {Wright}, {Xu}, {Zavagno}, {Zemcov}, {Zhang}, \& {Zonca}}]{GriffinA&A2010}
{Griffin}, M.~J., {Abergel}, A., {Abreu}, A., {et~al.} 2010, A\&A, 518, L3

\bibitem[{{Groves} {et~al.}(2015){Groves}, {Schinnerer}, {Leroy}, {Galametz},
  {Walter}, {Bolatto}, {Hunt}, {Dale}, {Calzetti}, \&
  {Croxall}}]{GrovesApJ2015}
{Groves}, B.~A., {Schinnerer}, E., {Leroy}, A., {et~al.} 2015, ApJ, 799, 96

\bibitem[{{Hildebrand}(1983)}]{HildebrandQJRAS1983}
{Hildebrand}, R.~H. 1983, QJRAS, 24, 267

\bibitem[{{Hodge}(1983)}]{HodgePASP1983}
{Hodge}, P.~W. 1983, PASP, 95, 721

\bibitem[{{Hunt} {et~al.}(2019){Hunt}, {De Looze}, {Boquien}, {Nikutta},
  {Rossi}, {Bianchi}, {Dale}, {Granato}, {Kennicutt}, \& {Silva}}]{HuntA&A2019}
{Hunt}, L.~K., {De Looze}, I., {Boquien}, M., {et~al.} 2019, A\&A, 621, A51

\bibitem[{{Hunt} {et~al.}(2015{\natexlab{a}}){Hunt}, {Draine}, {Bianchi},
  {Gordon}, {Aniano}, {Calzetti}, {Dale}, {Helou}, {Hinz}, {Kennicutt},
  {Roussel}, {Wilson}, {Bolatto}, {Boquien}, {Croxall}, {Galametz}, {Gil de
  Paz}, {Koda}, {Mu{\~n}oz-Mateos}, {Sandstrom}, {Sauvage}, {Vigroux}, \&
  {Zibetti}}]{HuntA&A2015}
{Hunt}, L.~K., {Draine}, B.~T., {Bianchi}, S., {et~al.} 2015{\natexlab{a}},
  A\&A, 576, A33

\bibitem[{{Hunt} {et~al.}(2015{\natexlab{b}}){Hunt}, {Garc{\'{\i}}a-Burillo},
  {Casasola}, {Caselli}, {Combes}, {Henkel}, {Lundgren}, {Maiolino}, {Menten},
  {Testi}, \& {Weiss}}]{HuntA&A2015b}
{Hunt}, L.~K., {Garc{\'{\i}}a-Burillo}, S., {Casasola}, V., {et~al.}
  2015{\natexlab{b}}, A\&A, 583, A114

\bibitem[{{James} {et~al.}(2002){James}, {Dunne}, {Eales}, \&
  {Edmunds}}]{JamesMNRAS2002}
{James}, A., {Dunne}, L., {Eales}, S., \& {Edmunds}, M.~G. 2002, MNRAS, 335,
  753

\bibitem[{{Jenkins}(2009)}]{JenkinsApJ2009}
{Jenkins}, E.~B. 2009, ApJ, 700, 1299

\bibitem[{{Jensen} {et~al.}(2005){Jensen}, {Rachford}, \&
  {Snow}}]{JensenApJ2005}
{Jensen}, A.~G., {Rachford}, B.~L., \& {Snow}, T.~P. 2005, ApJ, 619, 891

\bibitem[{{Jones} {et~al.}(2013){Jones}, {Fanciullo}, {K{\"o}hler},
  {Verstraete}, {Guillet}, {Bocchio}, \& {Ysard}}]{JonesA&A2013}
{Jones}, A.~P., {Fanciullo}, L., {K{\"o}hler}, M., {et~al.} 2013, A\&A, 558,
  A62

\bibitem[{{Jones} {et~al.}(2017){Jones}, {K{\"o}hler}, {Ysard}, {Bocchio}, \&
  {Verstraete}}]{JonesA&A2017}
{Jones}, A.~P., {K{\"o}hler}, M., {Ysard}, N., {Bocchio}, M., \& {Verstraete},
  L. 2017, A\&A, 602, A46

\bibitem[{{K{\"o}hler} {et~al.}(2015){K{\"o}hler}, {Ysard}, \&
  {Jones}}]{KoehlerA&A2015}
{K{\"o}hler}, M., {Ysard}, N., \& {Jones}, A.~P. 2015, A\&A, 579, A15

\bibitem[{{Lenz} {et~al.}(2017){Lenz}, {Hensley}, \& {Dor{\'e}}}]{LenzApJ2017}
{Lenz}, D., {Hensley}, B.~S., \& {Dor{\'e}}, O. 2017, ApJ, 846, 38

\bibitem[{{Li} \& {Draine}(2001)}]{LiApJ2001}
{Li}, A. \& {Draine}, B.~T. 2001, ApJ, 554, 778

\bibitem[{{Liszt}(2014)}]{LisztApJ2014}
{Liszt}, H. 2014, ApJ, 780, 10

\bibitem[{{Magnelli} {et~al.}(2014){Magnelli}, {Lutz}, {Saintonge}, {Berta},
  {Santini}, {Symeonidis}, {Altieri}, {Andreani}, {Aussel}, {B{\'e}thermin},
  {Bock}, {Bongiovanni}, {Cepa}, {Cimatti}, {Conley}, {Daddi}, {Elbaz},
  {F{\"o}rster Schreiber}, {Genzel}, {Ivison}, {Le Floc'h}, {Magdis},
  {Maiolino}, {Nordon}, {Oliver}, {Page}, {P{\'e}rez Garc{\'{\i}}a},
  {Poglitsch}, {Popesso}, {Pozzi}, {Riguccini}, {Rodighiero}, {Rosario},
  {Roseboom}, {Sanchez-Portal}, {Scott}, {Sturm}, {Tacconi}, {Valtchanov},
  {Wang}, \& {Wuyts}}]{MagnelliA&A2014}
{Magnelli}, B., {Lutz}, D., {Saintonge}, A., {et~al.} 2014, A\&A, 561, A86

\bibitem[{{Magrini} {et~al.}(2011){Magrini}, {Bianchi}, {Corbelli}, {Cortese},
  {Hunt}, {Smith}, {Vlahakis}, {Davies}, {Bendo}, {Baes}, {Boselli}, {Clemens},
  {Casasola}, {de Looze}, {Fritz}, {Giovanardi}, {Grossi}, {Hughes}, {Madden},
  {Pappalardo}, {Pohlen}, {di Serego Alighieri}, \&
  {Verstappen}}]{MagriniA&A2011}
{Magrini}, L., {Bianchi}, S., {Corbelli}, E., {et~al.} 2011, A\&A, 535, A13

\bibitem[{{Mannucci} {et~al.}(2010){Mannucci}, {Cresci}, {Maiolino}, {Marconi},
  \& {Gnerucci}}]{MannucciMNRAS2010}
{Mannucci}, F., {Cresci}, G., {Maiolino}, R., {Marconi}, A., \& {Gnerucci}, A.
  2010, MNRAS, 408, 2115

\bibitem[{{Martin} {et~al.}(2012){Martin}, {Roy}, {Bontemps},
  {Miville-Desch{\^e}nes}, {Ade}, {Bock}, {Chapin}, {Devlin}, {Dicker},
  {Griffin}, {Gundersen}, {Halpern}, {Hargrave}, {Hughes}, {Klein}, {Marsden},
  {Mauskopf}, {Netterfield}, {Olmi}, {Patanchon}, {Rex}, {Scott}, {Semisch},
  {Truch}, {Tucker}, {Tucker}, {Viero}, \& {Wiebe}}]{MartinApJ2012}
{Martin}, P.~G., {Roy}, A., {Bontemps}, S., {et~al.} 2012, ApJ, 751, 28

\bibitem[{{Mathis} {et~al.}(1983){Mathis}, {Mezger}, \&
  {Panagia}}]{MathisA&A1983}
{Mathis}, J.~S., {Mezger}, P.~G., \& {Panagia}, N. 1983, A\&A, 128, 212

\bibitem[{{Mathis} {et~al.}(1977){Mathis}, {Rumpl}, \&
  {Nordsiek}}]{MathisApJ1977}
{Mathis}, J.~S., {Rumpl}, W., \& {Nordsiek}, K.~H. 1977, ApJ, 217, 425

\bibitem[{{Mattsson} \& {Andersen}(2012)}]{MattssonMNRAS2012}
{Mattsson}, L. \& {Andersen}, A.~C. 2012, MNRAS, 423, 38

\bibitem[{{Meyer} {et~al.}(1998){Meyer}, {Jura}, \& {Cardelli}}]{MeyerApJ1998}
{Meyer}, D.~M., {Jura}, M., \& {Cardelli}, J.~A. 1998, ApJ, 493, 222

\bibitem[{{Mezger} {et~al.}(1982){Mezger}, {Mathis}, \&
  {Panagia}}]{MezgerA&A1982}
{Mezger}, P.~G., {Mathis}, J.~S., \& {Panagia}, N. 1982, A\&A, 105, 372

\bibitem[{{Nersesian} {et~al.}(2019){Nersesian}, {Xilouris}, {Bianchi},
  {Galliano}, {Jones}, {Baes}, {Casasola}, {Cassar{\`a}}, {Clark}, {Davies},
  {Decleir}, {Dobbels}, {De Looze}, {De Vis}, {Fritz}, {Galametz}, {Madden},
  {Mosenkov}, {Tr{\v{c}}ka}, {Verstocken}, {Viaene}, \&
  {Lianou}}]{NersesianA&A2019}
{Nersesian}, A., {Xilouris}, E.~M., {Bianchi}, S., {et~al.} 2019, A\&A, 624,
  A80

\bibitem[{{Nguyen} {et~al.}(2018){Nguyen}, {Dawson}, {Miville-Desch{\^e}nes},
  {Tang}, {Li}, {Heiles}, {Murray}, {Stanimirovi{\'c}}, {Gibson},
  {McClure-Griffiths}, {Troland}, {Bronfman}, \& {Finger}}]{NguyenApJ2018}
{Nguyen}, H., {Dawson}, J.~R., {Miville-Desch{\^e}nes}, M.-A., {et~al.} 2018,
  ApJ, 862, 49

\bibitem[{{Pettini} \& {Pagel}(2004)}]{PettiniMNRAS2004}
{Pettini}, M. \& {Pagel}, B.~E.~J. 2004, MNRAS, 348, L59

\bibitem[{{Pilbratt} {et~al.}(2010){Pilbratt}, {Riedinger}, {Passvogel},
  {Crone}, {Doyle}, {Gageur}, {Heras}, {Jewell}, {Metcalfe}, {Ott}, \&
  {Schmidt}}]{PilbrattA&A2010}
{Pilbratt}, G.~L., {Riedinger}, J.~R., {Passvogel}, T., {et~al.} 2010, A\&A,
  518, L1

\bibitem[{{Pilyugin} {et~al.}(2003){Pilyugin}, {Ferrini}, \&
  {Shkvarun}}]{PilyuginA&A2003}
{Pilyugin}, L.~S., {Ferrini}, F., \& {Shkvarun}, R.~V. 2003, A\&A, 401, 557

\bibitem[{{Planck Collaboration Int. XVII}(2014)}]{PlanckIntermediateXVII}
{Planck Collaboration Int. XVII}. 2014, A\&A, 566, A55

\bibitem[{{Planck Collaboration Int. XXIX}(2016)}]{PlanckIntermediateXXIX}
{Planck Collaboration Int. XXIX}. 2016, A\&A, 586, A132

\bibitem[{{Planck Collaboration VIII}(2014)}]{Planck2013VIII}
{Planck Collaboration VIII}. 2014, A\&A, 571, A8

\bibitem[{{Planck Collaboration XI}(2014)}]{Planck2013XI}
{Planck Collaboration XI}. 2014, A\&A, 571, A11

\bibitem[{{Planck Collaboration XXI}(2011)}]{PlanckEarlyXXI}
{Planck Collaboration XXI}. 2011, \aap, 536, A21

\bibitem[{{Planck Collaboration XXIV}(2011)}]{PlanckEarlyXXIV}
{Planck Collaboration XXIV}. 2011, A\&A, 536, A24

\bibitem[{{Pohlen} {et~al.}(2010){Pohlen}, {Cortese}, {Smith}, {Eales},
  {Boselli}, {Bendo}, {Gomez}, {Papageorgiou}, {Auld}, {Baes}, {Bock},
  {Bradford}, {Buat}, {Castro-Rodriguez}, {Chanial}, {Charlot}, {Ciesla},
  {Clements}, {Cooray}, {Cormier}, {Dwek}, {Eales}, {Elbaz}, {Galametz},
  {Galliano}, {Gear}, {Glenn}, {Griffin}, {Hony}, {Isaak}, {Levenson}, {Lu},
  {Madden}, {O'Halloran}, {Okumura}, {Oliver}, {Page}, {Panuzzo}, {Parkin},
  {Perez-Fournon}, {Rangwala}, {Rigby}, {Roussel}, {Rykala}, {Sacchi},
  {Sauvage}, {Schulz}, {Schirm}, {Smith}, {Spinoglio}, {Stevens}, {Srinivasan},
  {Symeonidis}, {Trichas}, {Vaccari}, {Vigroux}, {Wilson}, {Wozniak}, {Wright},
  \& {Zeilinger}}]{PohlenA&A2010}
{Pohlen}, M., {Cortese}, L., {Smith}, M.~W.~L., {et~al.} 2010, A\&A, 518, L72

\bibitem[{{Przybilla} {et~al.}(2008){Przybilla}, {Nieva}, \&
  {Butler}}]{PrzybillaApJ2008}
{Przybilla}, N., {Nieva}, M.-F., \& {Butler}, K. 2008, ApJ, 688, L103

\bibitem[{{Reach} {et~al.}(1995){Reach}, {Dwek}, {Fixsen}, {Hewagama},
  {Mather}, {Shafer}, {Banday}, {Bennett}, {Cheng}, {Eplee}, {Leisawitz},
  {Lubin}, {Read}, {Rosen}, {Shuman}, {Smoot}, {Sodroski}, \&
  {Wright}}]{ReachApJ1995}
{Reach}, W.~T., {Dwek}, E., {Fixsen}, D.~J., {et~al.} 1995, ApJ, 451, 188

\bibitem[{{R{\'e}my-Ruyer} {et~al.}(2014){R{\'e}my-Ruyer}, {Madden},
  {Galliano}, {Galametz}, {Takeuchi}, {Asano}, {Zhukovska}, {Lebouteiller},
  {Cormier}, {Jones}, {Bocchio}, {Baes}, {Bendo}, {Boquien}, {Boselli},
  {DeLooze}, {Doublier-Pritchard}, {Hughes}, {Karczewski}, \&
  {Spinoglio}}]{RemyRuyerA&A2014}
{R{\'e}my-Ruyer}, A., {Madden}, S.~C., {Galliano}, F., {et~al.} 2014, A\&A,
  563, A31

\bibitem[{{Ritchey} {et~al.}(2018){Ritchey}, {Federman}, \&
  {Lambert}}]{RitcheyApJS2018}
{Ritchey}, A.~M., {Federman}, S.~R., \& {Lambert}, D.~L. 2018, ApJS, 236, 36

\bibitem[{{Santini} {et~al.}(2014){Santini}, {Maiolino}, {Magnelli}, {Lutz},
  {Lamastra}, {Li Causi}, {Eales}, {Andreani}, {Berta}, {Buat}, {Cooray},
  {Cresci}, {Daddi}, {Farrah}, {Fontana}, {Franceschini}, {Genzel}, {Granato},
  {Grazian}, {Le Floc'h}, {Magdis}, {Magliocchetti}, {Mannucci}, {Menci},
  {Nordon}, {Oliver}, {Popesso}, {Pozzi}, {Riguccini}, {Rodighiero}, {Rosario},
  {Salvato}, {Scott}, {Silva}, {Tacconi}, {Viero}, {Wang}, {Wuyts}, \&
  {Xu}}]{SantiniA&A2014}
{Santini}, P., {Maiolino}, R., {Magnelli}, B., {et~al.} 2014, A\&, 562, A30

\bibitem[{{Schruba} {et~al.}(2012){Schruba}, {Leroy}, {Walter}, {Bigiel},
  {Brinks}, {de Blok}, {Kramer}, {Rosolowsky}, {Sandstrom}, {Schuster},
  {Usero}, {Weiss}, \& {Wiesemeyer}}]{SchrubaAJ2012}
{Schruba}, A., {Leroy}, A.~K., {Walter}, F., {et~al.} 2012, AJ, 143, 138

\bibitem[{{Siebenmorgen} {et~al.}(2014){Siebenmorgen}, {Voshchinnikov}, \&
  {Bagnulo}}]{SiebenmorgenA&A2014}
{Siebenmorgen}, R., {Voshchinnikov}, N.~V., \& {Bagnulo}, S. 2014, A\&A, 561,
  A82

\bibitem[{{Smith} {et~al.}(2019){Smith}, {Clark}, {De Looze}, {Lamperti},
  {Saintonge}, {Wilson}, {Accurso}, {Brinks}, {Bureau}, \&
  {Chung}}]{SmithMNRAS2019}
{Smith}, M. W.~L., {Clark}, C. J.~R., {De Looze}, I., {et~al.} 2019, MNRAS,
  486, 4166

\bibitem[{{Symeonidis} {et~al.}(2013){Symeonidis}, {Vaccari}, {Berta}, {Page},
  {Lutz}, {Arumugam}, {Aussel}, {Bock}, {Boselli}, {Buat}, {Capak}, {Clements},
  {Conley}, {Conversi}, {Cooray}, {Dowell}, {Farrah}, {Franceschini},
  {Giovannoli}, {Glenn}, {Griffin}, {Hatziminaoglou}, {Hwang}, {Ibar},
  {Ilbert}, {Ivison}, {Le Floc'h}, {Lilly}, {Kartaltepe}, {Magnelli}, {Magdis},
  {Marchetti}, {Nguyen}, {Nordon}, {O'Halloran}, {Oliver}, {Omont},
  {Papageorgiou}, {Patel}, {Pearson}, {P{\'e}rez-Fournon}, {Pohlen}, {Popesso},
  {Pozzi}, {Rigopoulou}, {Riguccini}, {Rosario}, {Roseboom}, {Rowan-Robinson},
  {Salvato}, {Schulz}, {Scott}, {Seymour}, {Shupe}, {Smith}, {Valtchanov},
  {Wang}, {Xu}, {Zemcov}, \& {Wuyts}}]{SymeonidisMNRAS2013}
{Symeonidis}, M., {Vaccari}, M., {Berta}, S., {et~al.} 2013, MNRAS, 431, 2317

\bibitem[{{Tremonti} {et~al.}(2004){Tremonti}, {Heckman}, {Kauffmann},
  {Brinchmann}, {Charlot}, {White}, {Seibert}, {Peng}, {Schlegel}, {Uomoto},
  {Fukugita}, \& {Brinkmann}}]{TremontiApJ2004}
{Tremonti}, C.~A., {Heckman}, T.~M., {Kauffmann}, G., {et~al.} 2004, ApJ, 613,
  898

\bibitem[{{Utomo} {et~al.}(2019){Utomo}, {Chiang}, {Leroy}, {Sandstrom}, \&
  {Chastenet}}]{UtomoApJ2019}
{Utomo}, D., {Chiang}, I.~D., {Leroy}, A.~K., {Sandstrom}, K.~M., \&
  {Chastenet}, J. 2019, \apj, 874, 141

\bibitem[{{van der Kruit}(1990)}]{VanDerKruitProc1990}
{van der Kruit}, P.~C. 1990, in The Milky Way as a Galaxy, ed. I.~{King},
  G.~{Gilmore}, \& P.~C. {van der Kruit}, 331--346

\bibitem[{{V{\'{\i}}lchez} {et~al.}(2019){V{\'{\i}}lchez}, {Rela{\~n}o},
  {Kennicutt}, {De Looze}, {Moll{\'a}}, \& {Galametz}}]{VilchezMNRAS2019}
{V{\'{\i}}lchez}, J.~M., {Rela{\~n}o}, M., {Kennicutt}, R., {et~al.} 2019, MN,
  483, 4968

\bibitem[{{Wolfire} {et~al.}(2010){Wolfire}, {Hollenbach}, \&
  {McKee}}]{WolfireApJ2010}
{Wolfire}, M.~G., {Hollenbach}, D., \& {McKee}, C.~F. 2010, ApJ, 716, 1191

\bibitem[{{Wright} {et~al.}(1991){Wright}, {Mather}, {Bennett}, {Cheng},
  {Shafer}, {Fixsen}, {Eplee}, {Isaacman}, {Read}, {Boggess}, {Gulkis},
  {Hauser}, {Janssen}, {Kelsall}, {Lubin}, {Meyer}, {Moseley}, {Murdock},
  {Silverberg}, {Smoot}, {Weiss}, \& {Wilkinson}}]{WrightApJ1991}
{Wright}, E.~L., {Mather}, J.~C., {Bennett}, C.~L., {et~al.} 1991, ApJ, 381,
  200

\bibitem[{{Ysard} {et~al.}(2012){Ysard}, {Juvela}, {Demyk}, {Guillet},
  {Abergel}, {Bernard}, {Malinen}, {M{\'e}ny}, {Montier}, \&
  {Paradis}}]{YsardA&A2012}
{Ysard}, N., {Juvela}, M., {Demyk}, K., {et~al.} 2012, \aap, 542, A21

\bibitem[{{Zubko} {et~al.}(2004){Zubko}, {Dwek}, \& {Arendt}}]{ZubkoApJS2004}
{Zubko}, V., {Dwek}, E., \& {Arendt}, R.~G. 2004, ApJS, 152, 211

\end{thebibliography}

\begin{appendix}
\section{Analogies with C16}
\label{app:c16}

Using Equations~\ref{eq:emy}, ~\ref{eq:tau}, ~\ref{eq:kappa}, and \ref{eq:mwscale},
 we can write
\begin{equation}
\kappa_\nu=\frac{d^2}{ (M_\ion{H}{i} + M_\mathrm{H_2}) \left(\frac{D/H}{O/H}\right)_\mathrm{MW} \, O/H} \, \frac{F_\nu}{B_\nu(T_\mathrm{d})}.
\label{eq:kours}
\end{equation}
When we neglect the (minor) contribution of an additional component of dust at a warmer temperature, the derivation of $\kappa_\nu$ in C16 (their Eq. 10) is equivalent to Eq.~\ref{eq:kours}, provided that
\begin{equation}
\left(\frac{D/H}{O/H}\right)_\mathrm{MW} = \frac{\xi \, \varepsilon_\mathrm{d} \, f_{Z_\odot}\,  \delta_\mathrm{O}}{(O/H)_\odot}.
\label{eq:c16scale}
\end{equation}
The parameters in Eq.~\ref{eq:c16scale} are $\xi$, a correction to account for the number of elements in the ISM other than \ion{H}{i} and H$_2$ (i.e. He -mainly- and metals); $\varepsilon_\mathrm{d}$, a constant dust-to-metal mass fraction; $f_{Z_\odot}$ and $(O/H)_\odot$, the solar metal mass fraction and oxygen abundance, respectively; $\delta_\mathrm{O}$, a correction of the gas oxygen abundance to take the oxygen atoms into account that are depleted into dust \citep[for details, see C16;][]{ClarkMNRAS2019}. 

For the values adopted by C16, this is $((D/H)/(O/H))_\mathrm{MW} = 25 \pm 6$; for the updated quantities in \citet{ClarkMNRAS2019}, this becomes $20 \pm 9$, which is closer to the value we used here (Sect.~\ref{sec:mw}).
\end{appendix}

\end{document}